\documentclass{aa}  

\usepackage{graphicx}
\usepackage{txfonts}

\usepackage{natbib}
\bibpunct{(}{)}{;}{a}{}{,} 
\def\epeak{E$_{\rm peak}$}
\def\epi{E$_{\rm pi}$}
\def\epo{E$_{\rm po}$}
\def\eiso{E$_{\rm iso}$}

\def\eer{\epi\ -- \eiso\ {\rm relation}}
\def\eep{\epi\ -- \eiso\ {\rm plane}}
\def\eeb{\epi\ -- \eiso\ {\rm boundary}}
\def\t9{T$_{90}$}

\def\vvm{V/V$_{\rm max}$}
\def\ntot{286}
\def\nout{33} 
\def\noutz{four} 
\def\noutnoz{29} 
\def\nother{253} 
\def\nz{43}
\def\nnoz{243}
\def\nzok{34}
\def\nlat{nine}
\def\zmin{0.34}
\def\zmax{4.35}
\begin{document}

   \title{The \epeak -- \eiso\ relation revisited with Fermi GRBs}
   \subtitle{Resolving a long-standing debate?}

   \author{V. Heussaff
          \inst{1,2}
          \and
          J-L. Atteia\inst{1,2}
          \and
          Y. Zolnierowski\inst{3}
          }

   \institute{Universit\'e de Toulouse; UPS-OMP; IRAP; Toulouse, France -- \email{vincent.heussaff@irap.omp.eu}
         \and
	CNRS; IRAP; 14, avenue Edouard Belin, F-31400 Toulouse, France
         \and
	LAPP, Universit\'e de Savoie, CNRS/IN2P3, 9 chemin de Bellevue, BP 110, 74941 Annecy-le-Vieux, France\\
          \\
             }

   \date{}

 
  \abstract
  {}
   {We used a sample of GRBs detected by Fermi and Swift to reanalyze the correlation discovered by Amati et al. (2002, A\&A, 390, 81) between \epi , the peak energy of the prompt GRB emission, and \eiso , the energy released by the GRB assuming isotropic emission. This correlation has been disputed by various authors, and our aim is to assess whether it is an intrinsic GRB property or the consequence of selection effects.}
   {We constructed a sample of Fermi GRBs with homogeneous selection criteria, and we studied their distribution in the \eep . Our sample is made of \nz\ GRBs with a redshift and \nnoz\ GRBs without a redshift. We show that GRBs with a redshift follow a broad \eer , while GRBs without a redshift show several outliers. We use these samples to discuss the impact of selection effects associated with GRB detection and with redshift measurement.  
}
   {We find that the \eer\ is partly due to intrinsic GRB properties and partly due to selection effects. The lower right boundary of the \eer\ stems from a true lack of luminous GRBs with low \epi. In contrast, the upper left boundary is attributed to selection effects acting against the detection GRBs with low \eiso\ and large \epi\ that appear to have a lower signal-to-noise ratio. In addition, we demonstrate that GRBs with and without a redshift follow different distributions in the \eep . GRBs with a redshift are concentrated near the lower right boundary of the \eer . This suggests that it is easier to measure the redshift of GRBs close to the lower \eeb\ and that GRBs with a redshift follow the Amati relation better than the general population. In this context, we attribute the controversy about the reality of the Amati relation to the complex nature of this relation resulting from the combination of a true physical boundary  and biases favoring the detection and the measurement of the redshift of GRBs located close to this boundary. 
}
   {}


   \maketitle
%

\section{Introduction}
\label{sec_intro}

\begin{figure}
\centering
\includegraphics[width=9cm]{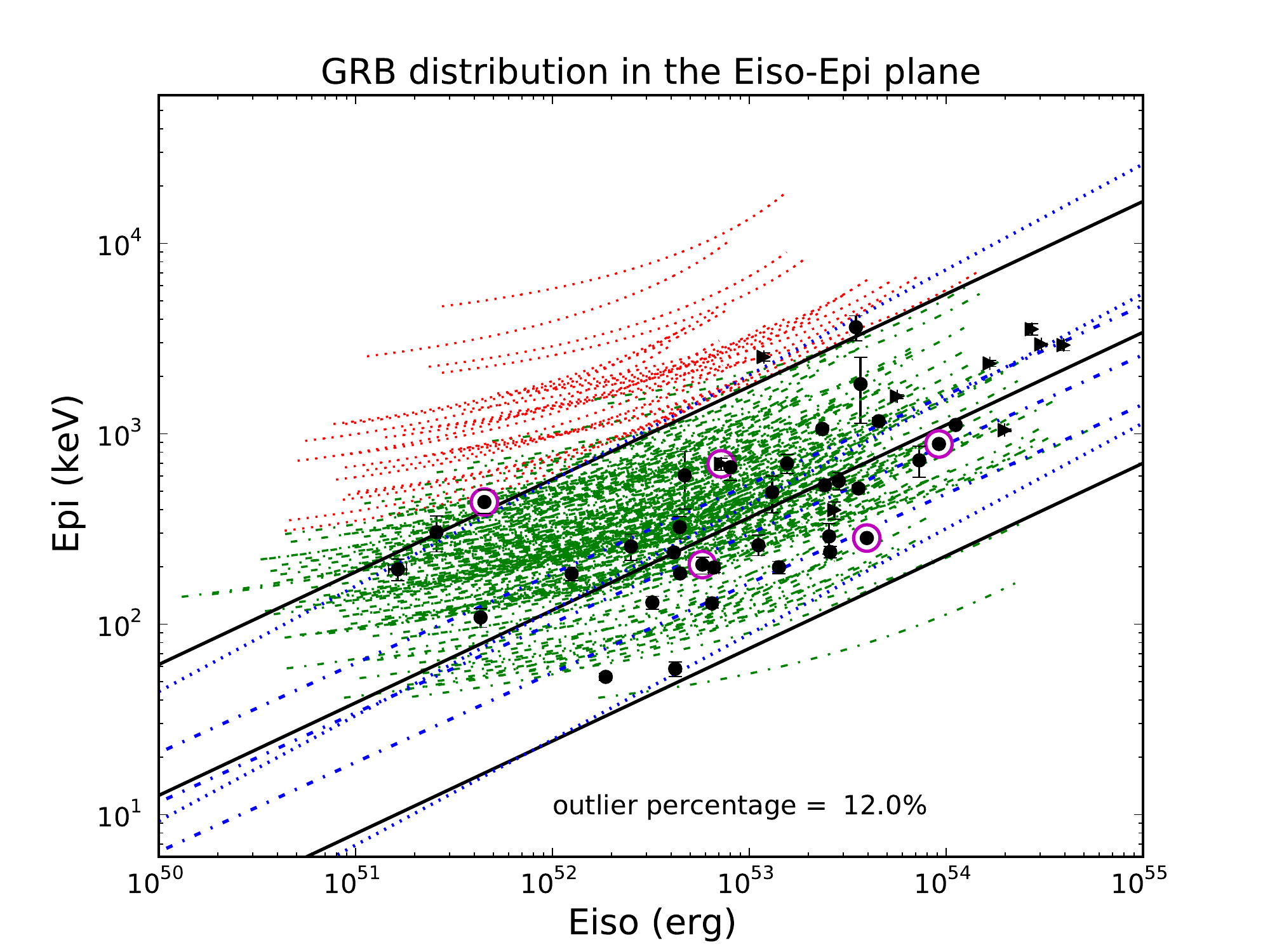}
\caption{Fermi GRBs in the plane \epi\ --\eiso . The black symbols show \nz\ GRBs with a redshift and the black solid lines show the corresponding best fit \eer\  with its two sigma limits. Black circles are \nzok\ GRBs detected simultaneously by Fermi/GBM and Swift/BAT. Black triangles are \nlat\ GRBs detected simultaneously by Fermi/GBM  and another positioning instrument (usually Fermi/LAT). The purple circles indicate GRBs whose redshift has been measured with the emission lines of the host galaxy. The green and red lines show the possible positions of GRBs without a redshift when they are placed at distances ranging from z=\zmin\ (leftmost point) to z=\zmax\ (rightmost point). Red lines indicate GRBs that are beyond the two sigma limit of the \eer\ at any redshift. The dotted blue lines show the best fit \eer\ of \cite{Nava2012} with its two sigma limits. The dashed blue lines show the best fit \eer\ of \cite{Amati2006} with its two sigma limits
}
\label{fig_outliers}
\end{figure}

Gamma-ray bursts (GRBs) are powerful cosmic explosions that are visible up to high redshifts (z $>5$) \citep{Lamb2000,Boer2006,Bloom2009,Vedrenne2009}. GRBs have two phases: a very bright flash of high-energy radiation usually lasting several seconds (sometimes also seen in visible light), called the prompt emission, followed by a long-lasting decaying afterglow, visible at longer wavelengths during several hours to months. This behavior is understood as the manifestation of a transient relativistic jet pointing towards the Earth, with the prompt emission being released inside the jet and the afterglow resulting from the interaction of the jet with the surrounding medium. 
The prompt emission is usually recorded with wide-field gamma-ray imagers or spectrometers and the afterglow with follow-up telescopes in space and on the ground. One essential element of GRB observations is the measurement of their redshift, which is done with visible or near infrared spectrographs on large ground-based telescopes. Most often, this measurement relies on the identification of absorption lines from the host galaxy in the spectrum of the afterglow, while it is bright enough. After measuring the redshift of a GRB one can correct from cosmological effects and infer its \textit{intrinsic} properties.

A surprising result obtained by \cite{Amati2002} was the discovery of a strong correlation between \epi , the peak energy of the $\nu \rm{F} \nu$ spectrum of the prompt emission at the source, and \eiso , the energy released by a GRB assuming isotropic emission. This correlation was first observed on a sample of ten GRBs with a redshift and well measured spectral parameters \citep{Amati2002}. After this discovery, the validity of this \epi\ -- \eiso\ relation (often called Amati relation) has been widely discussed, with conflicting results \citep[e.g.][]{Band2005, Ghirlanda2005, Nakar2005, Sakamoto2006, Butler2007, Cabrera2007, Schaefer2007, Butler2009, Firmani2009, Krimm2009, Butler2010, Shahmoradi2011, Collazzi2012a, Kocevski2012, Goldstein2012b}. Despite these discussions, the \epi\ -- \eiso\ relation was quickly proposed as a tool to infer GRB redshifts \citep[][]{Atteia2003}, to constrain the physics of the prompt emission \citep[][]{Eichler2004, Rees2005} and the geometry of GRB jets \citep[][]{Lamb2005,Toma2005}, and to \textit{standardize} GRBs, in much the same way as the Phillips relation was used to standardize SNe Ia \citep[][]{Phillips1993}. This last approach aims at constructing standard candles with GRBs and at using them to constrain the cosmological parameters \cite[e.g.][]{Dai2004, Friedman2005, Ghirlanda2006}. 

One difficulty of the study of the \eer\ is the need to measure both the peak energy of the $\nu \rm{F} \nu$ spectrum (called \epo\ in the observer's frame, and \epi\ in the source frame) and the redshift. The simultaneous operation of Swift and Fermi since July 2008 \citep{Gehrels2004, Atwood2009, Meegan2009} has provided a new sample of bursts whose parameters are available, allowing exploring the \eer\  over a larger parameter space than it was previously possible \citep[see also][]{Amati2010}. This paper discusses the distribution in the plane \epi\ -- \eiso\ of \nz\ GRBs with a redshift and \nnoz\ GRBs without a redshift, all events having \epo\ measured with Fermi/GBM. The GRB samples are presented in section \ref{sec_sample}. The distribution in the plane \epi\ -- \eiso\ is discussed in section \ref{sec_eer}. The impact of selection effects on this distribution is addressed in sections \ref{sec_outliers} and \ref{sec_redshift}, this last section addressing more specifically the selection effects associated with the measurement of redshift. The main results of our study and their interpretation are summarized in section \ref{sec_discussion}.

In all this paper we use a flat $\Lambda {\rm CDM}$ cosmology with $H_0 = 70$ km.s$^{-1}$.Mpc$^{-1}$ and $\Omega_\Lambda = 0.7$, since these values allow the direct comparison of our results with those of \cite{Nava2012}. Calculations of luminosity distances are done using the analytical approximation of \cite{Wickramasinghe2010} that has an accuracy better than 0.3\% in the redshift range used in this paper.


\section{GRB sample}
\label{sec_sample}

We select events with well measured spectral parameters from the Fermi GRB spectral catalog of \cite{Goldstein2012a} (see also \cite{Paciesas2012}) that provides the spectral parameters of 482 GRBs between GRB080714086 and GRB100709602 (about two years of data). We parametrize GRB spectra with the Band function \citep{Band1993} that consists of two smoothly connected power laws. Following standard naming, we call $\alpha$ the photon spectral index of the low-energy power law ($\alpha > -2$), and $\beta$ the photon spectral index of the high-energy power law ($\beta < -2$).  The $\nu F \nu$ spectrum peaks at the energy \epeak, near the junction of the two power laws. The selection is made with the application of the following cuts: 

\begin{itemize}
\item First, we make a selection on the duration. We consider GRBs with \t9 between 2 and 1000 seconds. This criterion excludes short GRBs (\t9 <2 sec), and very long GRBs that are superimposed on a varying background and whose \epeak\ is difficult to measure accurately.

\item Second, we require accurate spectral parameters. We exclude GRBs with one or more spectral parameters missing. We exclude GRBs with an error on alpha (the low-energy index of the Band function) larger than 0.4. We exclude a few GRBs with $\alpha < -2.0$, and GRBs with $\beta > \alpha$ because such values suggest a confusion between fitting parameters. We exclude GRBs with large errors on \epo , defined by a ratio of the 90\% upper limit to the 90\% lower limit larger than 3. We have less stringent constraints on beta (the high-energy index of the Band function) since we have checked that the position of GRBs in the \eep\ changes very little with beta. When the error on $\beta$ in the catalog is lacking or larger than 1.0, we assign to $\beta$ the classical value of -2.3 and we give no error. In a few cases, the high energy spectral index in the Fermi catalog is $> -2$, and the catalog gives the energy of a spectral break that is not \epo . In these cases we look for \epo\ in the GCN Circulars, and if we cannot find it, we simply remove the burst from the sample. This procedure suppresses one GRB with a redshift and ten GRBs without a redshift.  

\end{itemize}

267 bursts pass the cuts, 24 with a measured redshift. This is a small number, so we extend the sample of GRBs with a redshift with the addition of 19 Fermi GRBs found in GCN circulars that respect the same selection criteria. In the end, we have two samples: \nz\ GRBs with a redshift, and \nnoz\ GRBs without one. GRBs in each sample are listed in tables \ref{tab_zlist} and \ref{tab_nozlist}. The redshifts span the range [\zmin --\zmax ], with a median value \~{z} =1.71, smaller than the median redshift of Swift GRBs (\~{z} = 2.14), measured by \cite{Jakobsson2012}.

\section{The \epi\ -- \eiso\ relation of Fermi GRBs}
\label{sec_eer}
\subsection{GRBs with a redshift}
\label{sub_eer_redshift}
Figure~\ref{fig_outliers} shows with black symbols the locations of \nz\ Fermi GRBs with a redshift in the \eep . These bursts show a clear correlation between \epi\ and \eiso , as for the previous missions. The coefficient of correlation is 0.70, and the best fit correlation is ${\rm E}_{\rm pi} = 118 ~{\rm E}_{52}^{0.486}~ {\rm keV}$, where E$_{52}$ is the GRB isotropic energy in units of $10^{52}$ ergs (the best fit correlation is obtained by weighting each point quadratically by its error on \epi ). This will be the definition of the \eer\ in the rest of this paper. 

This relation is close to the one found by \cite{Gruber2012a} from a sample of Fermi GRBs (${\rm E}_{\rm pi} = 120\ {\rm E}_{52}^{0.55}~ {\rm keV}$), and by \cite{Nava2012} for a larger sample including GRBs from Fermi and other missions (${\rm E}_{\rm pi} = 119\ {\rm E}_{52}^{0.554}~ {\rm keV}$). If we freeze the slope to its best-fit value (0.486), we can compute the vertical standard deviation of data points around the fit: sigma~=~0.34 (if we measure the standard deviation perpendicular to the best fit line, we get sigma~=~0.21). The two-sigma limits are shown on Figure~\ref{fig_outliers}, they are defined by ${\rm E}_{\rm pi}= 24\ {\rm E}_{52}^{0.486}~ {\rm keV}$ for the lower limit, and ${\rm E}_{\rm pi} = 575\ {\rm E}_{52}^{0.486}~ {\rm keV}$, for the upper limit. 
Figure~\ref{fig_outliers} evidences a broadening of the two sigma confidence region with respect to the relation of \cite{Amati2006} for pre-Fermi GRBs that had a vertical sigma~=~0.21. This broadening was already noted by \cite{Nava2012} and by \cite{Gruber2012a}, who measure sigma~=~0.34 along the vertical axis, equivalent to the value found here. 

\subsection{GRBs without a redshift}
\label{sub_eer_noredshift}
The green and red dashed lines in Figure~\ref{fig_outliers} represent the possible locations of GRBs without a redshift in the \eep , considering that their redshift may vary between \zmin\ and \zmax , the extreme values measured for GRBs with a redshift. We plot in green GRBs lying within the two sigma limits of the \eer\ for some redshifts, and in red the bursts that remain outside these boundaries at any redshift. 

For most of the redshift range [\zmin --\zmax], the green and red lines are nearly parallel to the \eer , allowing assessing the position of GRBs without redshifts with respect to this relation. 
First, we note a clear asymmetry between GRBs above and below the best fit \eer . While GRBs with a redshift are more or less symmetric with respect to the best fit \eer\ (22 above and 21 below), the situation is markedly different for GRBs without a redshift. 72\% of them lie entirely above the best fit \eer , 22\% cross it, and  6\% only lie entirely below it. This asymmetry is accompanied by a lack of events in the lower right corner of the plot. Since GRBs in this region have large \eiso\ and low \epi , they should have plenty of photons and we cannot attribute their deficit to selection effects. We conclude that GRBs with a large \eiso\ and a low \epi\ are intrinsically rare. 

We finally note that \noutnoz\ GRBs are located outside the two sigma limits of the \eer\ at all redshifts, all of them lie \textit{above} the two sigma upper limit of the \eer .
This is larger than the number of six GRBs expected on a purely statistical basis (assuming a gaussian shape of the distribution around the \eer ). It is difficult to infer the exact significance of this excess, because our definition of the two-sigma limits is rather arbitrary and the uncertainties on \epeak\ make difficult evaluating the true number of outliers. Despite these uncertainties, it is not possible to move more than a few outliers within the two-sigma limits, and we conclude safely that Fermi GRBs without a redshift include a significant fraction of outliers to the \eer\ .

At this point we are facing two results: on one hand GRBs with a redshift seem to follow a broad \eer , while on the other hand the overall GRB population shows a fraction of outliers of 12\% in a region of the plane where GRBs are difficult to detect, suggesting a much broader distribution in the \eep . 
The following sections discuss how the observed \epi\ -- \eiso\ distribution is affected by selection effects. In section \ref{sec_outliers} we discuss selection effects on GRB detection, while in section \ref{sec_redshift} we discuss selection effects on the measurement of the redshift. 

\begin{figure}
\centering
\includegraphics[width=9cm]{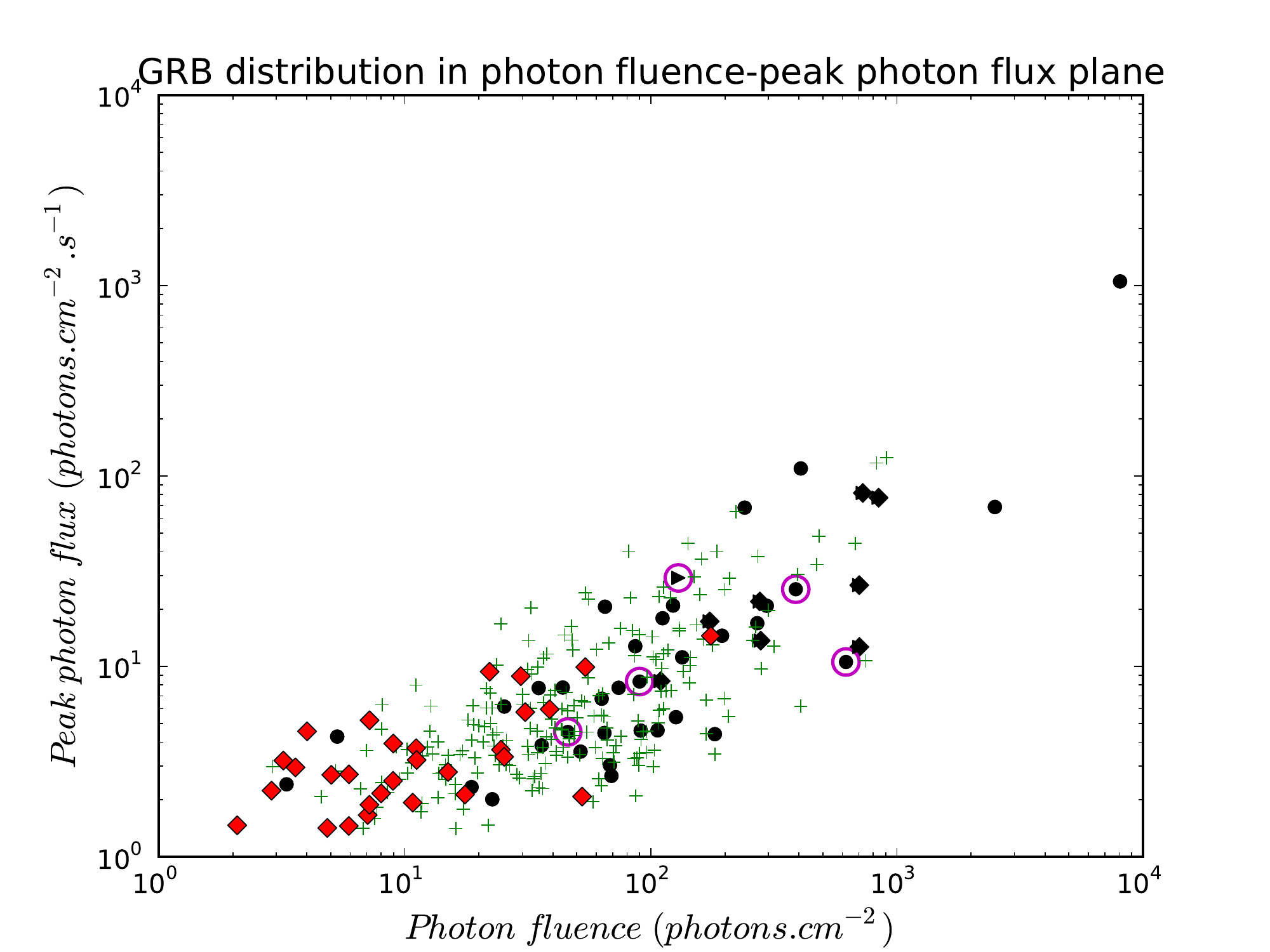}
\caption{Distribution of three classes of GRBs (see Figure~\ref{fig_outliers}) in a photon fluence -- peak photon flux plane. Several outliers to the \eer\ (in red) are GRBs at the limit of detection, with low photon fluence and peak photon flux.
}
\label{fig_brightness}
\end{figure}

\section{Study of the outliers}
\label{sec_outliers}

\begin{table}
\caption{Comparison of the observed properties of \nout\ GRBs incompatible with \eer\ at all redshifts (\textit{outliers}) with \nother\ GRBs that are within two sigma of the \eer\ (\textit{others}). The KS significance gives the probability that a parameter has the same distribution for outliers and for other GRBs.}
\centering
\begin{tabular}{|lccc|}
\hline
Parameter  & median & median & KS \\
\textit{(unit)}  & (outliers) & (others) & significance  \\
\hline
\t9  & 22.5  & 29.7 & 0.17 \\
\textit{(s) }&  &  & \\
Alpha  & -0.88  & -0.89 & 0.93 \\
 &  &  & \\
Beta  & -2.30  & -2.30 & 0.49 \\
 &  &  & \\
\epo  & 715  & 151 & 3.5 $10^{-16}$ \\
\textit{(keV)} &  &  & \\
Photon Fluence  & 10.8  & 53.7 & 6.2 $10^{-7}$ \\
\textit{(ph cm$^{-2}$)} &  &  & \\
Energy Fluence  & 28  & 74 & 1.4 $10^{-3}$ \\
\textit{($10^{-7}$ erg cm$^{-2}$)} &  &  & \\
Peak Photon Flux  & 3.2  & 5.5 & 2.5 $10^{-3}$ \\
\textit{(ph cm$^{-2}$ s$^{-1}$)} &  &  & \\
\hline
\end{tabular}
\label{tab_KS_outliers}
\end{table}

\begin{figure*}
\centering
\includegraphics[width=6cm]{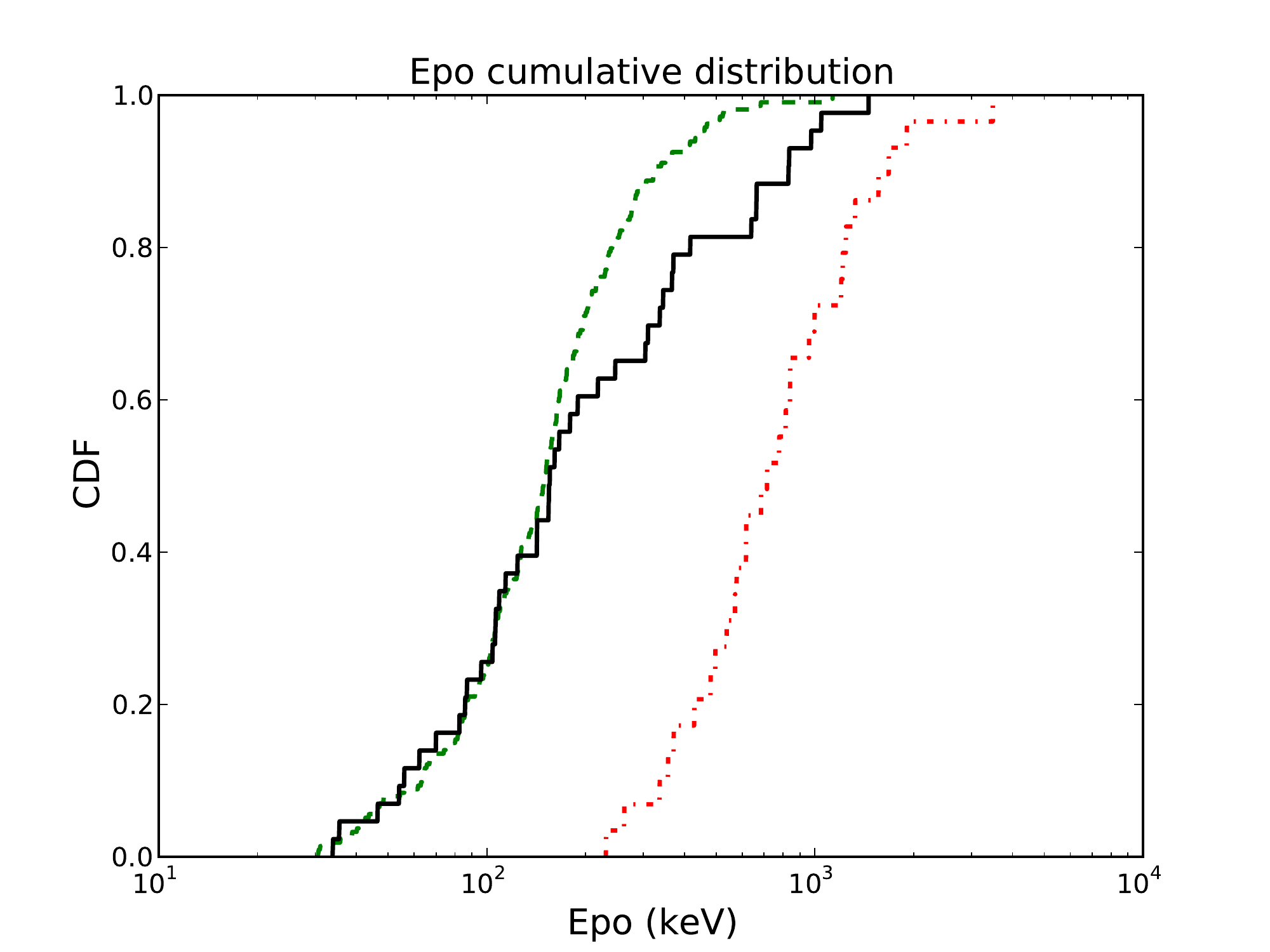}
\includegraphics[width=6cm]{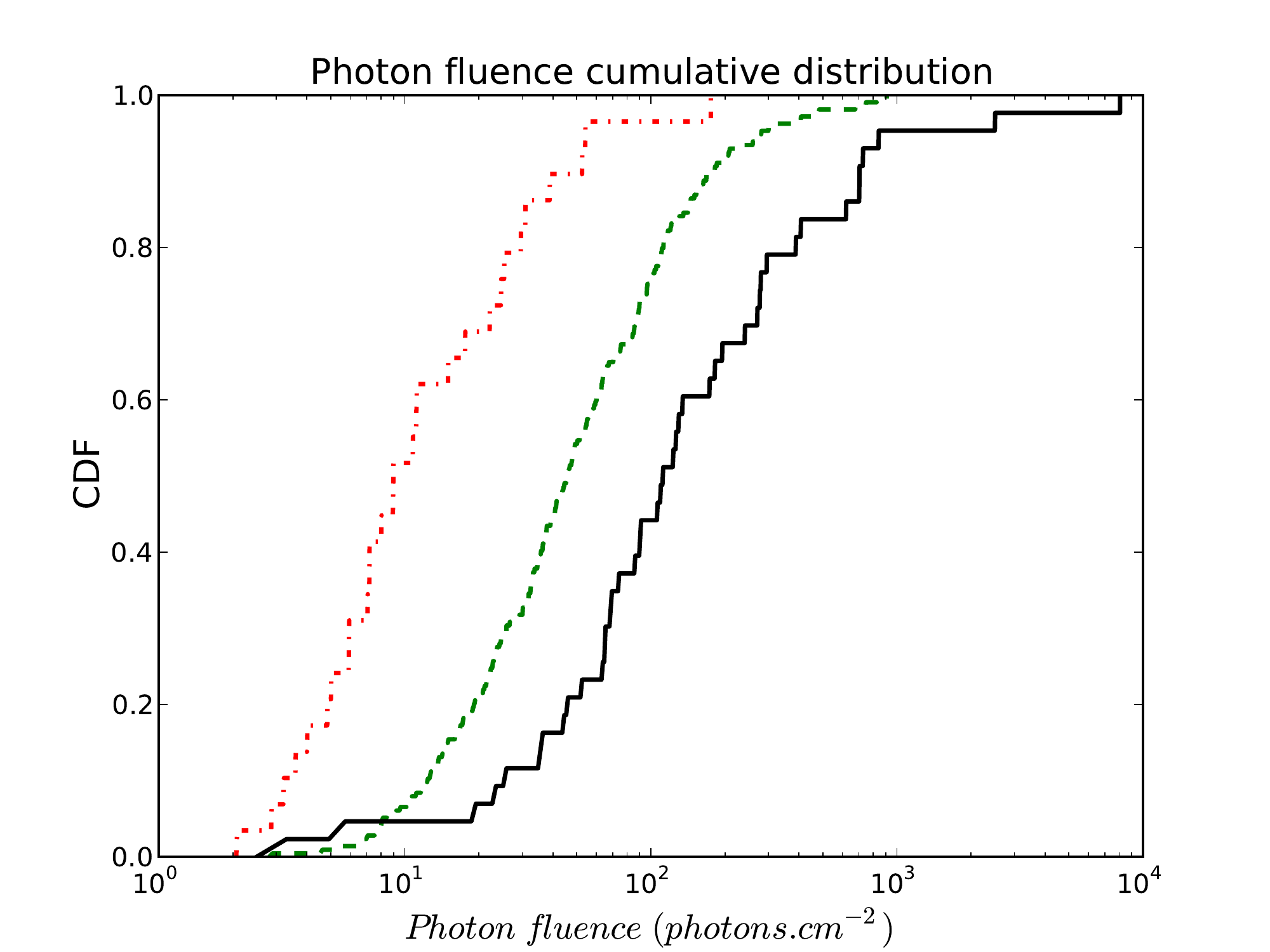}
\includegraphics[width=6cm]{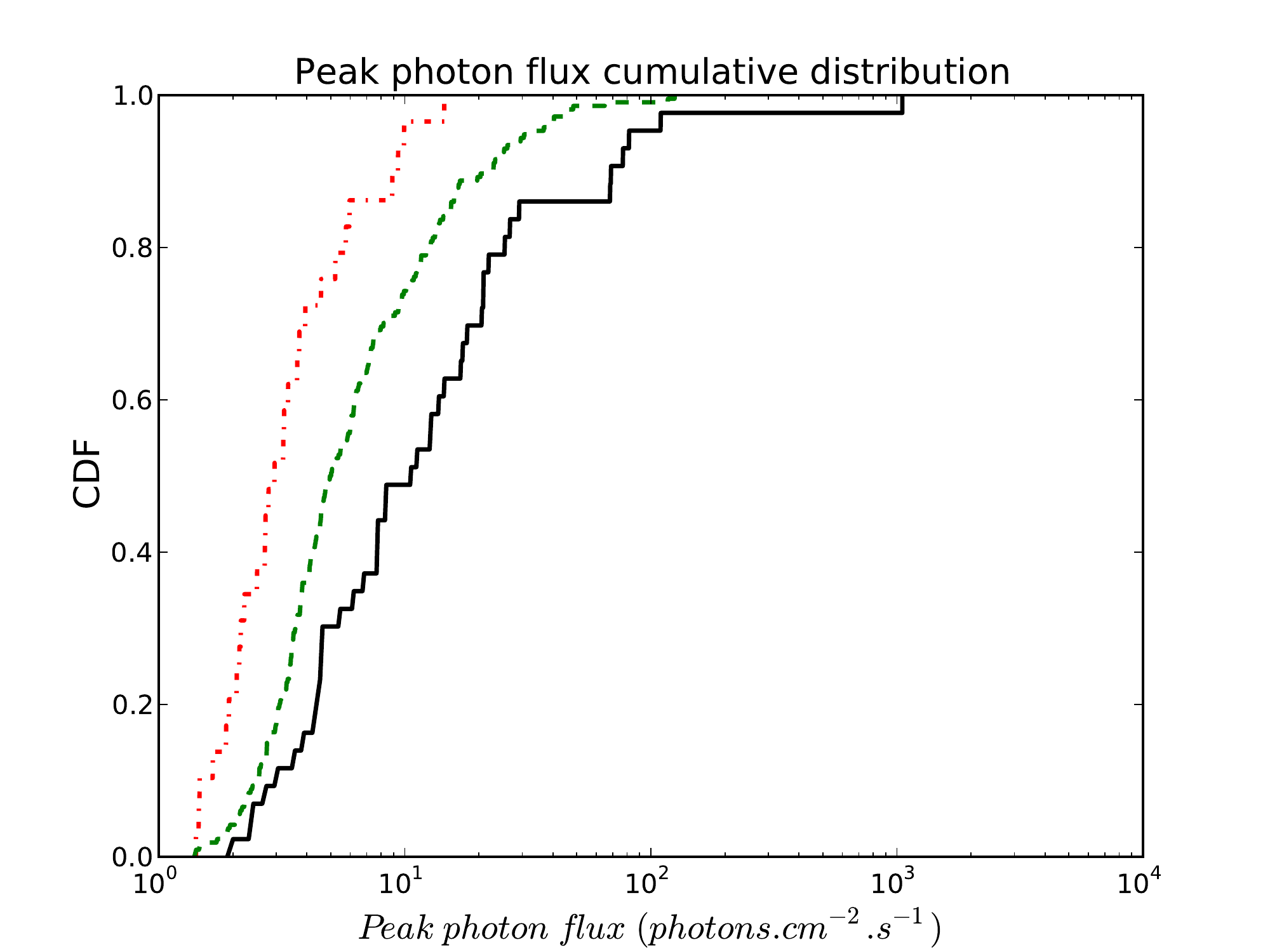}
\caption{Cumulative distribution function (CDF) of 3 GRB parameters:  \epo , 
photon fluence, and peak photon flux for the three classes of Fermi GRBs defined in Figure~\ref{fig_outliers}. Black solid line: \nz\ GRBs with a redshift; green dashed line: \nother\ GRBs with or without a redshift that may be compatible with the \epi\ -- \eiso\ relation at some redshift; red dash-dotted line: \noutnoz\ GRBs without a redshift that are incompatible with the \epi\ -- \eiso\ relation at any redshift (outliers). This figure emphasizes the larger \epo\ and smaller photon fluence of outliers.
}
\label{fig_distributions}
\end{figure*}

In the rest of this paper, we call \textit{outliers} GRBs located above the two-sigma limit of the \eer. There are \nout\ outliers, \noutz\ GRBs with a redshift, and \noutnoz\ GRBs without a redshift, they are shown in boldface in tables \ref{tab_zlist} and \ref{tab_nozlist}. In this section we compare the properties of the \nout\ outliers, with those of the remaining \nother\ GRBs, in order to understand whether they are affected by significant selection effects. 
Table \ref{tab_KS_outliers} shows the results of Kolmogorov-Smirnov (KS) tests comparing the duration \t9, spectral parameters, and brightness between the two classes. Outliers differ from GRBs that follow the \eer\ by their \epo\ and their brightness: on average, outliers a have larger \epo\ (a median \epo\ of 715 keV vs 151 keV) and a smaller fluence (a median fluence of 28 $10^{-7}$ erg cm$^{-2}$ vs 74 $10^{-7}$ erg cm$^{-2}$). These differences are clearly visible in Figure~\ref{fig_distributions} that shows the cumulative distribution functions (CDF) of these parameters: on average outliers have an observed \epeak\ four times larger and a photon fluence five times smaller than other GRBs. 

The main consequence is that outliers are much more difficult to detect \citep[see][for a discussion of GRB sensitivity of different detectors]{Band2003}. In fact, it is difficult to assess the true extension of the GRB population above the \eer\ because GRBs in this region have fewer photons as the direct consequence of their large \epo . The fraction of outliers found here is thus a lower limit of the true fraction of outliers. We conclude that,  above the \eer , the observed GRB distribution is currently limited by the sensitivity of gamma-ray detectors. 

Similar conclusions were reached by \cite{Nakar2005}, \cite{Goldstein2010}, and \cite{Shahmoradi2011} based on BATSE GRBs, by \cite{Butler2007} based on Swift GRBs and by \cite{Collazzi2012a} based on various GRB samples from pre-Fermi missions, and by \cite{Kocevski2012} based on a synthetic GRB population. Other authors have stressed the role of selection effects, however they consider that they have little impact on the \eer . \cite{Ghirlanda2008} show that the existence of an \epo -- fluence correlation could impact the \eer\ of Swift GRBs and that pre-Swift GRBs also could be affected by unknown selection effects, and \cite{Nava2008} states that, although selection effects are present, they do not determine the spectral-energy correlations. 
In order to clarify this debate, we now compare the distribution of GRBs with a redshift and GRBs without a redshift in the \eep .

\section{GRBs with a redshift}
\label{sec_redshift}

In order to compare GRBs with a redshift and GRBs without a redshift, we discuss the differences between the two classes in terms of their observed properties in section \ref{sub_observed}, and in terms of their intrinsic properties in section \ref{sub_intrinsic}. We limit the sample of GRBs with a redshift to the bursts that have the same selection criteria as GRBs without a redshift, that is to a sample of \nzok\ GRBs detected \textit{simultaneously} by Fermi/GBM (for the spectral parameters) and Swift/BAT (for the localization). This excludes \nlat\ GRBs detected and localized by Fermi/LAT, IPN, or INTEGRAL that were outside the field-of-view of Swift/BAT at the time of the burst. Including these events in the sample of GRBs with a redshift would artificially increase the fraction of very bright GRBs, biasing the comparison between Fermi GRBs with and without a redshift, as explained in section \ref{sub_LAT} below.  

\subsection{Observed properties}
\label{sub_observed}
Table \ref{tab_KS_redok} compares the properties of GRBs with a redshift with the population of GRBs with no redshift. 
GRBs with a redshift differ from the general population by their larger brightness (fluence and peak flux). This is not surprising since GRBs that are brighter in gamma-rays are also expected to be brighter in visible (e.g. \cite{Wanderman2010}). 
The larger average fluence and peak flux of GRBs with a redshift is explained in part by a smaller distance (the median redshift of our sample is \~{z} = 1.71, compared \~{z} = 2.14 for Swift GRBs.) and in part by a greater ease to measure the redshift of brighter GRBs. This effect however cannot create the \eer\ that is an intrinsic property of GRBs. It is thus necessary to check whether the measurement of the redshift biases the intrinsic properties of GRBs, and more specifically their distribution in the \eep.

\begin{table}
\caption{Comparison of the observed properties of \nzok\ GRBs with a redshift and \nnoz\ GRBs without a redshift. The KS significance gives the probability that a parameter has the same distribution for GRBs with a redshift and GRBs without a redshift.}
\centering
\begin{tabular}{|lccc|}
\hline
Parameter  & median & median & KS \\
\textit{(unit)}  & (with z) & (without z) & significance  \\
\hline
\t9  & 43.5  & 28.2  & 0.10 \\
\textit{(s)} &  &  & \\
Alpha  & -0.92  & -0.87  & 0.58 \\
 &  &  & \\
Beta  & -2.30  & -2.30  & 0.58 \\
 &  &  & \\
\epo   & 142  & 162  & 0.44 \\
\textit{(keV)}  &  &  & \\
Photon Fluence   & 88  & 40  & 7.3 $10^{-4}$ \\
\textit{(ph cm$^{-2}$)} &  &  & \\
Energy Fluence   & 108  & 58  & 1.4 $10^{-3}$ \\
\textit{($10^{-7}~erg cm^{-2}$)} &  &  & \\
Peak Photon Flux  & 7.7  & 4.7  & 6.4 $10^{-2}$ \\
\textit{(ph cm$^{-2}$ s$^{-1}$)} &  &  & \\
\hline
\end{tabular}
\label{tab_KS_redok}
\end{table}

\subsection{Intrinsic properties}
\label{sub_intrinsic}
\begin{figure*}
\centering
\includegraphics[width=9cm]{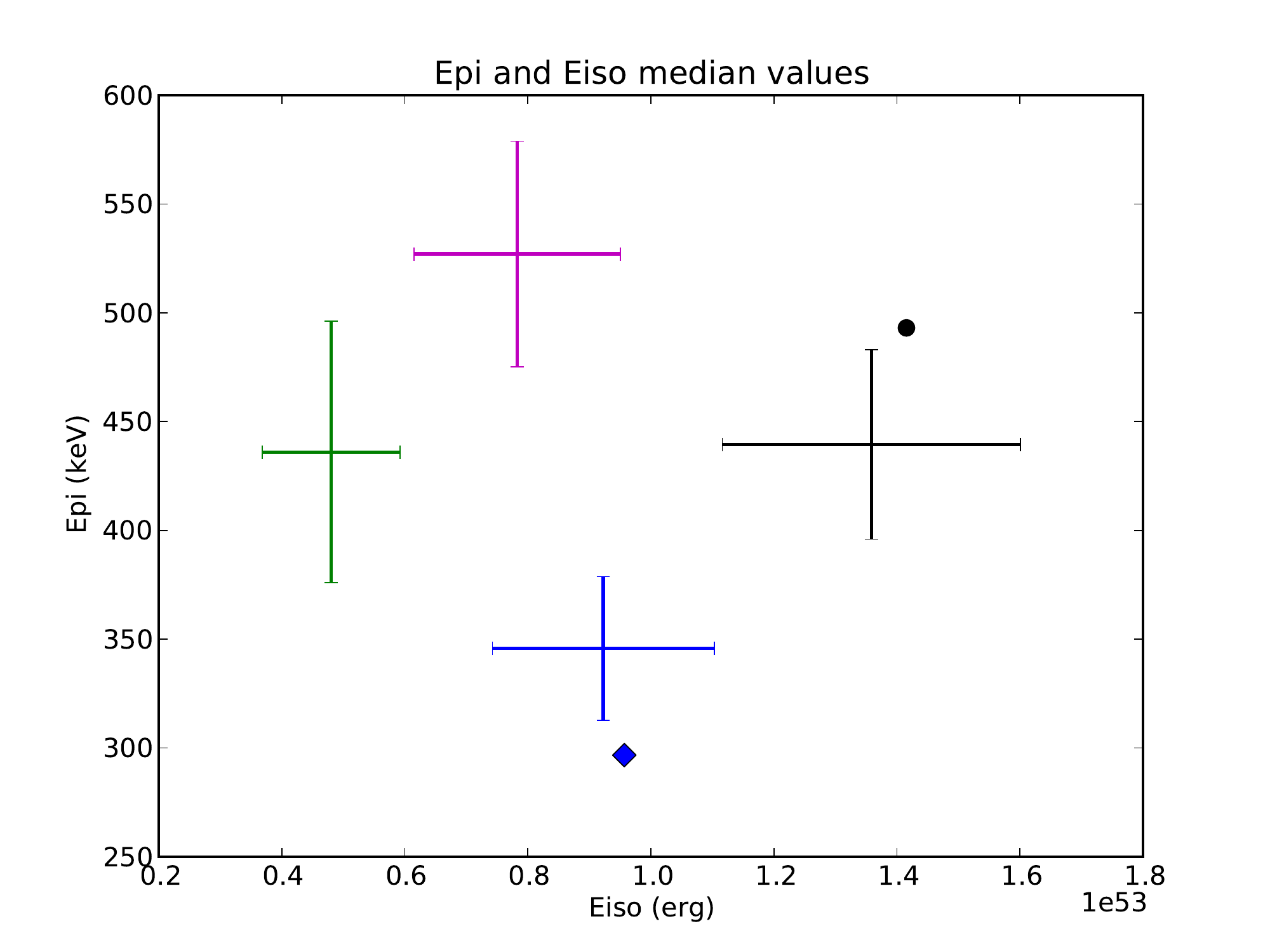}
\includegraphics[width=9cm]{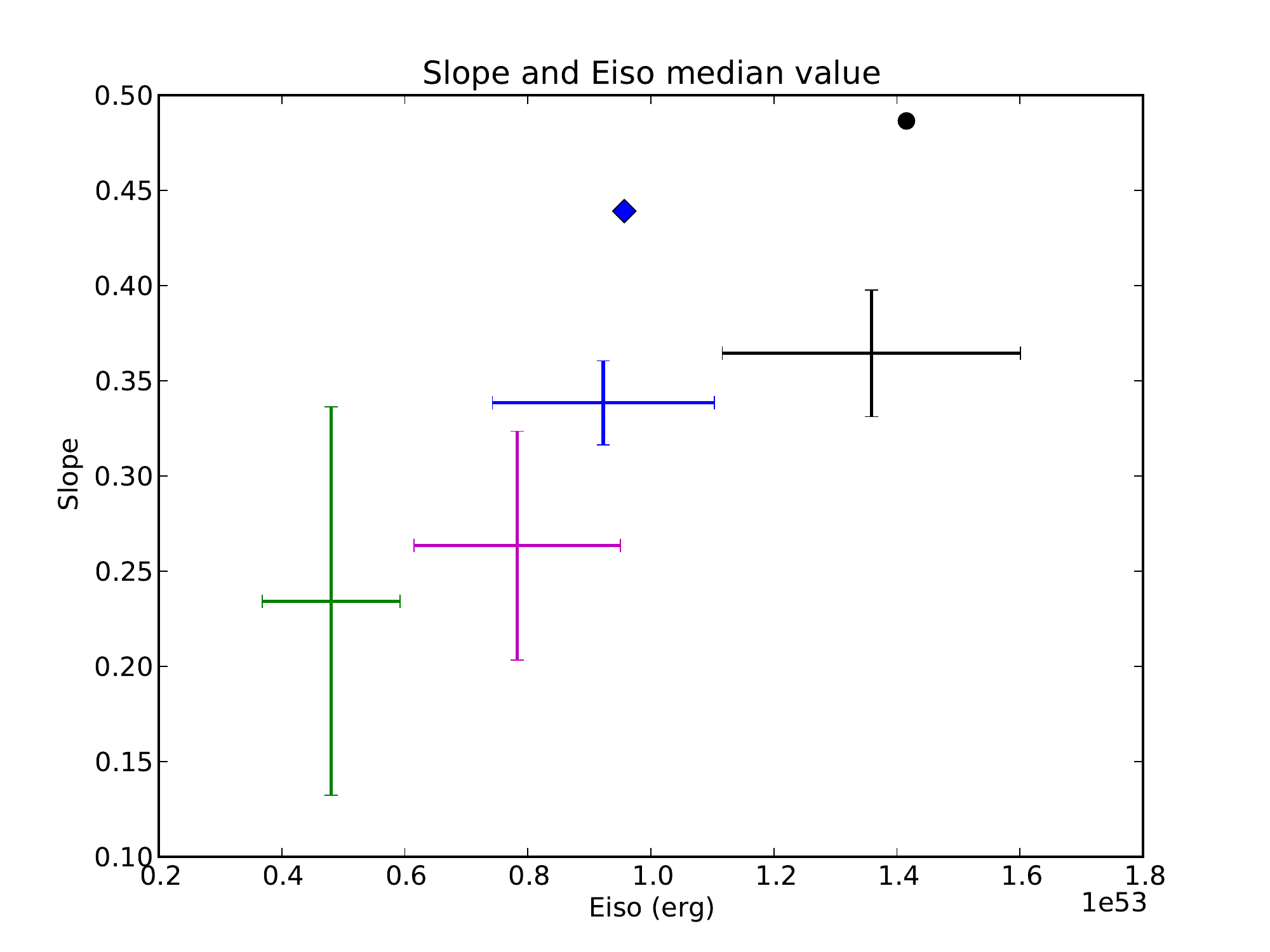}
\caption{Comparison of simulated GRB populations. \textit{Left panel.} \epi\ as a function of \eiso. \textit{Right panel.} Slope of the \eer\ as a function of \eiso. Green: Fermi GRBs with no redshift not localized by Swift/BAT; Purple: Fermi GRBs with no redshift localized by Swift/BAT; Blue: Fermi GRBs with a redshift localized by Swift/BAT; Black: Fermi GRBs with a redshift, irrespective of the satellite used for the localization. The crosses show the median values and the corresponding error, measured from 1000 simulated samples. The point symbols show the values measured for two actual samples: in blue \nzok\ GRBs studied in section \ref{sub_intrinsic}, and in black the full sample of \nz\ GRBs with a redshift. 
}
\label{fig_mediane}
\end{figure*}

In order to check whether the measurement of the redshift impacts the distribution of GRBs in the \eep , we use the following procedure: from the sample of \nnoz\ GRBs without a redshift, we randomly select \nzok\ GRBs, we assign them the redshifts of the \nzok\ GRBs with measured redshifts, and we compute their \epi , \eiso , and the slope of the best fit \eer . We repeat this procedure 1000 times and we compare the median values of \epi , \eiso , and the slope of the \eer\ with those of GRBs with redshifts. We apply the same procedure to the subsample of \nzok\ GRBs with a redshift, in order to understand how the redshift redistribution impacts the results. For sake of completeness, we study two additional distributions: 25 Fermi GRBs localized by Swift without a redshift, and the full sample of \nz\ GRBs with a redshift. The results are summarized in Figure~\ref{fig_mediane}. The crosses show the median \eiso , \epi , the median slope of the \eer\ and the one-sigma error on these values, for the four simulated GRBs samples. The blue and black symbols show the values measured for GRBs with a redshift (excluding or not the GRBs detected by Fermi/LAT).

These simulations bring the following results: 
\begin{itemize}
      \item The compatibility (within errors) of the green and purple crosses shows that the localization with Swift/BAT does not bias the distribution of \epi\ or \eiso .
      \item The comparison of the blue and green crosses shows that GRBs with a redshift have slightly larger \eiso\ (median = $9.2~10^{52}$ erg vs $4.8~10^{52}$ erg), similar \epi\ (median = 345 keV vs 440 keV), and slopes of the \eer\ that are compatible ($\sim0.3$). The measurement of the redshift thus favors GRBs that are more energetic, with large \eiso . 
      \item The comparison of the simulated populations (blue and black crosses) with the actual GRBs (blue and black symbols) shows that the redshift redistribution builds simulated populations that have more or less the same \eiso\ and \epi\ as the original population, but a significantly smaller slope of the \eer . This difference can be attributed to the fact that, unlike real GRBs, simulated GRBs can occupy the empty region in the lower right corner of the \eep , decreasing the slope of the correlation. 
\end{itemize}

The differences pointed here between GRBs with a redshift and GRBs without a redshift concern average values of the parameters. They provide no information on the joint distribution in the plane \epi\ -- \eiso\ for the two groups. A look at Figure~\ref{fig_outliers} suggests different distributions of the two groups, with GRBs with a redshift gathered closer to the \eeb . To verify this hypothesis, we have divided our GRB sample (34 GRBs with a redshift plus 243 GRBs without a redshift) into four subsamples according to their distance to the \eeb\ (for GRBs with a redshift) or to their smallest distance to this line (for GRBs without a redshift). Since we have excluded the \nlat\ GRBs with a redshift detected by Fermi/LAT, each subsample includes eight or nine GRBs with a redshift. We have computed the fraction of GRBs with a redshift in each subsample, obtaining the following numbers  with increasing distance from the boundary: 9/44=0.20 ; 9/42=0.21 ;  8/68=0.12 ; and 8/123=0.065. These numbers clearly show that GRBs with a redshift do not follow the general GRB distribution in the plane \epi\ -- \eiso , being more concentrated near the \eeb . This is illustrated in Figure~\ref{fig_groupes} that shows with different colors the groups of GRBs at different distances from the \eeb . 

With this observation, we finally get a coherent picture of the origin of \eer . This correlation can be attributed to the combination of two effects: a physical boundary in the plane \epi\ -- \eiso , and biases favoring the detection and the measurement of redshift of GRBs located along this boundary. We have evidenced two biases acting in the same direction. First, for a given \eiso , GRBs far from the \eeb\ have fewer gamma-ray photons, making their detection and localization with gamma-ray detectors more difficult. This is confirmed with the computation of \vvm\ as a proxy for the signal-to-noise ratio.\footnote{\vvm\ is a number between 0 and 1 that represents the position of a GRB within its "volume of detectability". GRBs with a low signal-to-noise ratio that are close to the detection limit have \vvm $\lesssim 1$ ; in contrast GRBs with a large signal-to-noise ratio have \vvm $\approx 0$.} We find that the median \vvm\ increases with the distance from the \eeb , with  median(\vvm ) = 0.28 - 0.35 - 0.51 - 0.75 for the 4 groups of GRBs with a redshift defined in the previous paragraph. These numbers only concern a small fraction of GRBs but they clearly show that GRBs close to the \eeb\ have larger signal-to-noise and are more easily detected. 

The second bias is connected with the measurement of the redshift: we have found that the fraction of GRBs with a redshift increses close to the \eeb . This suggests the existence of a bias that makes easier the measurement of the redshift when a GRB is close to the \eeb . Since the capability to measure the redshift is closely linked with the optical brightness of the afterglow, we infer that there are factors enhancing the optical luminosity of GRBs close to the \eeb . The verification of this hypothesis requires a systematic study of the optical luminosity of GRB afterglows, which is beyond the scope of this paper. This bias contributes to further reduce the width of the distribution of GRBs with a redshift. As a consequence of these two biases, GRBs with a redshift appear preferentially along a narrow band above the \eeb , giving rise to correlation found by \cite{Amati2002}. 

\subsection{Bright GRBs detected with LAT and IPN}
\label{sub_LAT}
Figure~\ref{fig_outliers} shows that the 9 GRBs that were not localized by Swift (the black triangles) concentrate in the upper right of the \eer , increasing the significance of the correlation. In our interpretation, these very bright GRBs cannot have low \epi\ because of the \eeb . Since they brightness facilitates the measurement of their redshift, they play an important role in shaping the \eer . 

\begin{figure}
\centering
\includegraphics[width=9cm]{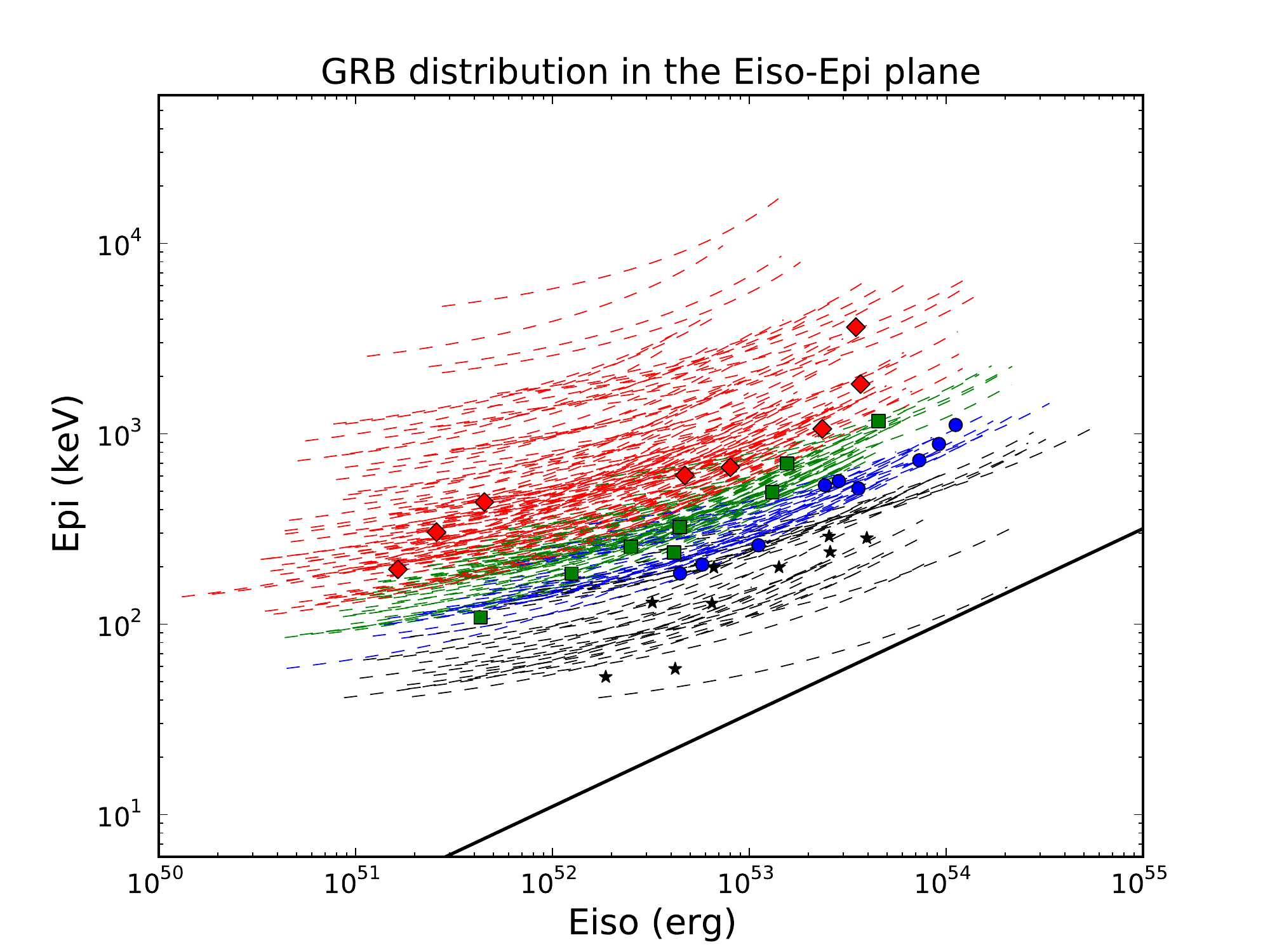}
\caption{GRBs classification in 4 groups according to their distance to the \eeb. The parameter used to separate the 4 groups is their vertical distance (for GRBs with a redshift) or their smallest vertical distance (for GRBs without a redshift) to the black line. The black line has a slope=0.486, assumed to be the slope of the \eeb . The position of the black line is arbitrary, it does not impact the groups as long as it stays below the GRB sample. 
}
\label{fig_groupes}
\end{figure}

This is also the case of GRB~130427A (the rightmost point in figures \ref{fig_brightness} and \ref{fig_groupes}) that is one of the few nearby energetic GRBs (like GRB 030329 detected by Hete-2 at z=0.17, \cite{Vanderspek2004}). GRB~130427A is at redshift z=0.34, the value of \vvm\ for this GRB, \vvm = 0.003, indicates that there is a 10\% chance observing this GRB so close among 30 similar bursts at higher redshift. Considering that Fermi has detected more than 1100 GRBs from its launch to the end of April 2013, the detection of a nearby energetic GRB like GRB~130427A is not unexpected. Interestingly, GRB~130427A lies perfectly along the Amati relation \citep{Amati2013}. This is compatible with our interpretation that highly energetic bursts cannot have low \epi\ because of the \eeb .

\section{Discussion}
\label{sec_discussion}

The results presented in this paper are based on the distribution of GRBs in the plane \epi\ -- \eiso . 
We discuss below some selection effects that may affect this distribution.
First, it is not possible to measure \epo\ outside the energy range of the detector. This restricts the range of measured \epo\ to [30 -- 1300] keV. These bounds prevent us to include X-Ray Flashes (soft GRBs with \epeak $<$ 30 keV) in our analysis but they do not bias the distribution of GRBs in the \eep\ and the conclusions of our analysis. 
Second, we have studied the impact of the errors on \epo\ and on the energy fluence S$_{\gamma}$ given in Table \ref{tab_nozlist}. If we systematically replace \epo\ with \epo $-\sigma_{\rm E_{po}}$ for all GRBs without a redshift, the number of outliers decreases from \noutnoz\ to 18. Similarly, if we change the fluence S$_{\gamma}$ with S$_{\gamma}-\sigma_{\rm S_\gamma}$ for all GRBs without a redshift, the number of outliers decreases from \noutnoz\ to 28. Thus, we cannot get rid of the outliers, even with this rather drastic method.

Two additional effects, acting in opposite directions, can affect the observed distribution in the \eep . 
First, GRBs in the upper left part of the plot are more difficult to detect because they have intrinsically fewer photons than GRBs closer to the Amati relation. Since this bias acts against the detection of outliers, it leads to underestimate the true fraction of outliers, as explained in section \ref{sec_outliers}. 
On the other hand, it is possible to 'create' outliers from GRBs following the Amati relation if the measurements of \epi\ or \eiso\ are strongly biased. 
One possibility is the "duration bias" discussed by \cite{Kocevski2013}. This bias results from the inability of count-limited instruments to detect the parts of GRBs with low count rates, artificially reducing \t9\ and \eiso\ of faint GRBs. 
The duration bias may also affect the measurement of \epeak\ since the brightest parts of faint GRBs are often those with the largest \epeak . 
Since Table \ref{tab_KS_outliers} shows that outliers have essentially the same duration distribution as the rest of GRBs, there is no hint in the data of a strong duration bias that could significantly alter our analysis. 
We have also checked the possible biases on \epeak\ with the measurement of the ratio RE = \epeak (at\ peak) / \epeak (average) for outliers and for other GRBs. We find that RE is broadly distributed with median(RE) = $0.74 \pm 0.30 $ for outliers, median(RE) = $1.14 \pm 0.28$ for other GRBs, and median(RE) = $1.06 \pm 0.28$ for GRBs with a redshift. These numbers show that outliers are only marginally affected by selection effects that could modify their \epeak .
Overall, we found no bias capable of 'creating' the observed outliers from GRBs that follow the \eer .

The main conclusion of this paper is that the \eer\ is due to the combination of two effects: a physical limit that prevents the existence of GRBs with a large \eiso\ and a low \epi\ (creating an \eeb ), and the sensitivity of gamma-ray and optical imagers that strongly favor the detection and localization of GRBs located close to the \eeb\ \citep[see also][]{Shahmoradi2011}. The combination of these two effects creates a band in the \eep\ where GRBs are more easily detected, and a thinner band where their redshift can be measured. We believe that this interpretation explains the apparently contradictory results obtained by people who have checked the validity of \eer\ either using large samples of GRBs without a redshift or only using GRBs with a redshift. The first have consistently found large fractions of outliers dismissing the relation \citep[e.g.][]{Band2005, Nakar2005, Butler2007, Butler2010, Shahmoradi2011, Collazzi2012a}, while the second usually confirmed its validity \citep[e.g.]{Ghirlanda2005, Sakamoto2006, Krimm2009, Nava2012}. The horizontal extent of the \eer\ depends on the GRB luminosity function, on the evolution of the GRB rate with redshift, and on the volume of detection for each type of GRB. We have checked that the leftmost part of the \eer\ contains mainly faint  GRBs with low \epi\ detected at low redshift (z<1), while the rightmost part is made of luminous GRBs with large \epi\ that can only be found in large volumes of observation (except GRB~130427A discussed in section \ref{sub_LAT}). Finally, we note that similar studies could be performed with other relations, like the \epi\ -- L$_{\rm iso}$ relation \citep{Yonetoku2004} or the \epi\ -- E$_\gamma$ relation \citep{Ghirlanda2004b}, to check their nature and their validity.

A discussion of the origin of the \eeb\ is beyond the scope of this paper. If it is a consequence of the dominant radiation mechanism at work in GRBs, as we expect, theoretical studies may help understanding the nature and the true location of this boundary \citep[see for instance][for explanations within the framework of different theoretical or phenomelogical models]{Eichler2004, Yamazaki2004, Rees2005, Thompson2007, Giannios2012, Mochkovitch2012, Beniamini2013, Shahmoradi2013}. The observation of an \eer\ in time-resolved spectra \citep[e.g.][]{Ghirlanda2010, Frontera2012, Basak2012b, Lu2012} may also provide insights into the origin of the \eeb . 
Finally, Since the \eer\ is not an intrinsic property of GRBs, it may be difficult to use it to standardize GRBs \citep{Ghirlanda2006, Meszaros2012}. This might nevertheless be possible if one finds a way of identifying GRBs close to the \eeb , thanks to their temporal properties (e.g. spectral lags) or to the characteristics of their afterglows. 

\begin{acknowledgements}
The authors gratefully acknowledge the use of the online Fermi GRB catalog (http://heasarc.gsfc.nasa.gov/W3Browse/fermi/ fermigbrst.html), the use of the GRB table compiled by J. Greiner (http://www.mpe.mpg.de/$\sim$jcg/grbgen.html), and the use of the table of GRB redshifts compiled by P\'all Jakobsson (http://raunvis.hi.is/$\sim$pja/GRBsample.html). YZ welcomes financial support from IRAP (UMR5277/CNRS/UPS).

\end{acknowledgements}

\bibliographystyle{aa} 
\bibliography{biblio-VH2012} 

\begin{appendix}
\section{Data tables}
\label{sec_tables}
The following tables provide the names and the spectral properties of the \ntot\ GRBs used in this study, extracted from the Fermi GRB catalog of \cite{Goldstein2012a}.

\begin{table*}
\caption{\nz\ GRBs with a redshift. The references for the redshift and spectral parameters are given at the bottom of the table. The 9 GRBs in italic have been localized by LAT, IPN or INTEGRAL, they are at the bright end of the GRB luminosity function. \noutz\ GRBs are outliers to the Amati relation, they are indicated in bold. The errors are given for a confidence level of 90 \%.}
\label{tab_zlist}
\centering
\begin{tabular}{|lccccccccc|}
\hline
GRB Name  & z \tablefootmark{*}  & $\alpha$  & $\alpha$ & $\beta$\tablefootmark{\dag}  & $\beta$  & \epo \tablefootmark{**}  &  \epo  & Fluence, S$_\gamma$  & Fluence \\
 &  &  & error &  & error  & (keV)  & error  & ($10^{-7}~erg~cm^{-2}$)  & error \\
\hline
GRB080804972    & 2.20 $^{a}$  & -0.52  & 0.09  & -1.90  & 0.08 & 218 $^{ba}$  & 25  & 91.3  & 1.1 \\
GRB080810549    & 3.35 $^{b}$  & -1.16  & 0.03  & -2.30  &  & 834 $^{ba}$  & 125  & 108.2  & 0.5 \\
GRB080905705    & 2.37 $^{c}$  & -0.85  & 0.23  & -2.28  & 0.62 & 179 $^{ba}$  & 60  & 29.1  & 0.4 \\
\textit{GRB080916009}    & 4.35 $^{d}$  & -1.08  & 0.01  & -2.15 & 0.07 & 662 $^{ba}$  & 45  & 602.7  & 0.7 \\
GRB080916406    & 0.69 $^{e}$  & -0.90  & 0.10  & -2.30  &  & 109 $^{bb}$  & 9  & 78.1  & 0.8 \\
\hline--
GRB081008832    & 1.97 $^{f}$  & -1.01  & 0.12  & -2.09  & 0.21 & 166 $^{ba}$  & 36  & 103.2  & 1.5 \\
GRB081121858    & 2.51 $^{g}$  & -0.43  & 0.13  & -2.09  & 0.09 & 161 $^{ba}$  & 17  & 152.7  & 2.2 \\
GRB081221681    & 2.26 $^{h}$  & -0.90  & 0.02  & -3.86  & 0.49 & 87 $^{ba}$  & 1  & 300.4  & 0.9 \\
GRB081222204    & 2.77 $^{i}$  & -0.86  & 0.05  & -2.31  & 0.12 & 142 $^{ba}$  & 10  & 118.9  & 1.0 \\
GRB090102122    & 1.55 $^{j}$  & -0.94  & 0.02  & -2.30  & & 417 $^{ba}$  & 18  & 279.1  & 0.6 \\
\hline
\textit{GRB090323002}    & 3.57 $^{k}$  & -1.29  & 0.01  & -2.45  & 0.18 & 639 $^{ba}$  & 41  & 1180.9  & 1.7 \\
\textit{\textbf{GRB090328401}}    & 0.74 $^{l}$  & -1.13  & 0.01  & -2.65  & 0.26 & 1458 $^{ba}$  & 74  & 420.2  & 0.7 \\
GRB090424592    & 0.54 $^{m}$  & -1.04  & 0.02  & -2.76  & 0.13 & 154 $^{ba}$  & 4  & 463.2  & 0.4 \\
GRB090516353    & 4.11 $^{n}$  & -1.52  & 0.05  & -2.30  & 0.27 & 142 $^{ba}$  & 26  & 172.2  & 0.6 \\
GRB090618353    & 0.54 $^{o}$  & -1.26  & 0.06  & -2.50  & 0.33 & 156 $^{bc}$  & 11  & 2684.5  & 4.3 \\
\hline
\textit{GRB090902462}    & 1.82 $^{p}$  & -1.00  & 0.00  & -2.30  &  & 1044 $^{ba}$  & 17  & 2217.9  & 3.2 \\
\textit{GRB090926181}    & 2.11 $^{q}$  & -0.85  & 0.01  & -2.39  & 0.04 & 336 $^{ba}$  & 6  & 1465.7  & 3.4 \\
GRB090926914    & 1.24 $^{r}$  & 0.13  & 0.12  & -3.25  & 0.34 & 82 $^{ba}$  & 3  & 107.9  & 1.5 \\
\textit{GRB091003191}    & 0.90 $^{s}$  & -1.07  & 0.02  & -2.22  & 0.11 & 366 $^{ba}$  & 27  & 233.2  & 0.8 \\
GRB091020900    & 1.71 $^{t}$  & -1.26  & 0.06  & -2.30  &  & 246 $^{ba}$  & 36  & 83.5  & 1.5 \\
\hline
GRB091127976    & 0.49 $^{u}$  & -1.26  & 0.07  & -2.22  & 0.02 & 35 $^{ba}$  & 2  & 207.3  & 0.4 \\
GRB091208410    & 1.06 $^{v}$  & -1.44  & 0.07  & -2.32  & 0.47 & 124 $^{bd}$  & 20  & 61.9  & 1.9 \\
\textit{GRB100414097}    & 1.37 $^{w}$  & -0.62  & 0.01  & -3.53  & 0.48 & 664 $^{ba}$  & 16  & 884.7  & 1.9 \\
GRB100615083    & 1.40 $^{x}$  & -1.24  & 0.08  & -2.27  & 0.12 & 86 $^{be}$  & 8  & 87.2  & 0.8 \\
GRB100728095    & 1.57 $^{y}$  & -0.75  & 0.01  & -3.04  & 0.57 & 344 $^{bf}$  & 9  & 1278.8  & 5.8 \\
\hline
GRB100728439    & 2.11 $^{z}$  & -0.80  & 0.20  & -2.20  & 0.20 & 104 $^{bg}$  & 14  & 33.4  & 0.6 \\
GRB100814160    & 1.44 $^{aa}$  & -0.64  & 0.14  & -2.02  & 0.12 & 106 $^{bh}$  & 14  & 149.4  & 0.9 \\
GRB100906576    & 1.73 $^{ab}$  & -1.34  & 0.08  & -1.98  & 0.07 & 106 $^{bi}$  & 18  & 232.9  & 0.6 \\
GRB101213451    & 0.41 $^{ac}$  & -1.10  & 0.07  & -2.35  & 0.72 & 310 $^{bj}$  & 49  & 74.0  & 1.0 \\
GRB101219686    & 0.55 $^{ad}$  & 0.33  & 0.36  & -2.12  & 0.12 & 70 $^{bk}$  & 8  & 39.9  & 0.5 \\
\hline
GRB110731465    & 2.83 $^{ae}$  & -0.80  & 0.03  & -2.98  & 0.30 & 304 $^{bl}$  & 13  & 228.9  & 0.6 \\
GRB111228657    & 0.71 $^{af}$  & -1.90  & 0.10  & -2.70  & 0.30 & 34 $^{bm}$  & 3  & 181.4  & 0.6 \\
GRB120119170    & 1.73 $^{ag}$  & -0.98  & 0.03  & -2.36  & 0.09 & 189 $^{bn}$  & 8  & 386.7  & 1.4 \\
GRB120326056    & 1.80 $^{ah}$  & -0.98  & 0.14  & -2.53  & 0.15 & 46 $^{bo}$  & 4  & 32.6  & 0.5 \\
\textit{GRB120711115}    & 1.41 $^{ai}$  & -0.94  & 0.01  & -2.40 & 0.04 & 973 $^{bp}$  & 35  & 1942.8  & 2.3 \\
\hline
\textit{GRB120716712}    & 2.49 $^{aj}$  & -1.00  & 0.08  & -2.08  & 0.07 & 114 $^{bq}$  & 12  & 146.9  & 0.5 \\
GRB120811649    & 2.67 $^{ak}$  & -0.71  & 0.20  & -2.30  & & 54 $^{br}$  & 3  & 34.5  & 2.1 \\
GRB120907017    & 0.97 $^{al}$  & -0.75  & 0.25  & -2.30  & & 154 $^{bs}$  & 33  & 8.1  & 0.4 \\
GRB120909070    & 3.93 $^{am}$  & -1.30  & 0.10  & -2.30  & & 370 $^{bt}$  & 140  & 98.5  & 1.5 \\
GRB121128212    & 2.20 $^{an}$  & -0.80  & 0.12  & -2.41  & 0.10 & 62 $^{bu}$  & 5  & 104.0  & 4.0 \\
\hline
GRB121211574    & 1.02 $^{ao}$  & -0.30  & 0.34  & -2.30  & & 96 $^{bv}$  & 13  & 4.8  & 0.5 \\
GRB130420313    & 1.30 $^{aq}$  & -1.00  & 0.13  & -2.30  & & 56 $^{bw}$  & 3  & 115.9  & 2.4 \\
GRB130427324    & 0.34 $^{ar}$  & -0.79  & 0.00  & -3.06  & 0.02 & 830 $^{bx}$  & 5  & 24620.0  & 12.4 \\
\hline
\end{tabular}\\
\tablefoot{
\tablefoottext{\dag}{when $\beta$ is unconstrained, we use $\beta = -2.3$}\\
\tablefoottext{*}{references for the redshift: $^{a)}$ \cite{Cucchiara2008a, Thoene2008} -- $^{b)}$ \cite{Burenin2008, Deugartepostigo2008, Prochaska2008} -- $^{c)}$ \cite{Vreeswijk2008} -- $^{d)}$ \cite{Greiner2009} -- $^{e)}$ \cite{Fynbo2008} -- $^{f)}$ \cite{Davanzo2008, Cucchiara2008b} -- $^{g)}$ \cite{Berger2008} -- $^{h)}$ \cite{Salvaterra2012} -- $^{i)}$ \cite{Cucchiara2008c, Graham2008} -- $^{j)}$ \cite{Cucchiara2009a, Deugartepostigo2009a} -- $^{k)}$ \cite{Chornock2009a} -- $^{l)}$ \cite{Cenko2009a} -- $^{m)}$ \cite{Chornock2009a, Wiersema2009a} - $^{n)}$ \cite{Deugartepostigo2009b, Rossi2009} -- $^{o)}$ \cite{Cenko2009b, Fatkhullin2009} -- $^{p)}$ \cite{Cucchiara2009b} -- $^{q)}$ \cite{Malesani2009} -- $^{r)}$ \cite{Fynbo2009} -- $^{s)}$ \cite{Cucchiara2009c} -- $^{t)}$ \cite{Xu2009} -- $^{u)}$ \cite{Cucchiara2009d, Thoene2009} -- $^{v)}$ \cite{Perley2009, Wiersema2009b} -- $^{w)}$ \cite{Cucchiara2010} -- $^{x)}$ \cite{Kruehler2013a} -- $^{y)}$ \cite{Kruehler2013b} -- $^{z)}$ \cite{Flores2010} -- $^{aa)}$ \cite{OMeara2010} -- $^{ab)}$ \cite{Tanvir2010} -- $^{ac)}$ \cite{Chornock2011} -- $^{ad)}$ \cite{Deugartepostigo2011, Sparre2011} -- $^{ae)}$ \cite{Tanvir2011} -- $^{af)}$ \cite{Cucchiara2011, Dittmann2011, Palazzi2011, Schulze2011, Xu2011} -- $^{ag)}$ \cite{Cucchiara2012, Milisavljevic2012, Pancoast2012} -- $^{ah)}$ \cite{Tello2012} -- $^{ai)}$ \cite{Tanvir2012a} -- $^{aj)}$ \cite{Delia2012, Greiner2012} -- $^{ak)}$ \cite{Fynbo2012, Thoene2012} -- $^{al)}$ \cite{SanchezRamirez2012} -- $^{am)}$ \cite{Hartoog2012} -- $^{an)}$ \cite{Tanvir2012b} -- $^{ao)}$ \cite{Perley2012} -- $^{ap)}$ \cite{Cucchiara2013} -- $^{aq)}$ \cite{Deugartepostigo2013} -- $^{ar)}$ \cite{Flores2010, Garnavich2013, Levan2013, Xu2013} }\\
}
\end{table*}

\begin{table*}
\tablefoot{
\tablefoottext{**}{references for the peak energy: $^{ba)}$ \cite{Goldstein2012a} -- $^{bb)}$ \cite{Bissaldi2008} -- $^{bc)}$ \cite{McBreen2009a} -- $^{bd)}$ \cite{McBreen2009b} -- $^{be)}$ \cite{Foley2010} -- $^{bf)}$ \cite{Vonkienlin2010a} -- $^{bg)}$ \cite{Vonkienlin2010b} -- $^{bh)}$ \cite{Vonkienlin2010c} -- $^{bi)}$ \cite{Gruber2010a} -- $^{bj)}$ \cite{Gruber2010b} -- $^{bk)}$ \cite{Vanderhorst2010} -- $^{bl)}$ \cite{Gruber2011} -- $^{bm)}$ \cite{Briggs2011} -- $^{bn)}$ \cite{Gruber2012b} -- $^{bo)}$ \cite{Collazzi2012b} -- $^{bp)}$ \cite{Gruber2012c} -- $^{bq)}$ \cite{Gruber2012d} -- $^{br)}$ \cite{Xiong2012} -- $^{bs)}$ \cite{Younes2012} -- $^{bt)}$ \cite{Chaplin2012} -- $^{bu)}$ \cite{Mcglynn2012} -- $^{bv)}$ \cite{Yu2012} -- $^{bw)}$ \cite{Xiong2013} -- $^{bx)}$ \cite{Vonkienlin2013}}\\
}
\end{table*}

\begin{table*}
\caption{\nnoz\ GRBs without a redshift. The spectral parameters are from \citep{Goldstein2012a}, except otherwise indicated.
\noutnoz\ GRBs are outliers to the Amati relation, they are indicated in italic. The errors are given for a confidence level of 90 \%.}
\label{tab_nozlist}
\centering
\begin{tabular}{|lcccccccc|}
\hline
GRB Name  & $\alpha$  & $\alpha$ & $\beta$\tablefootmark{a}  & $\beta$ & \epo \tablefootmark{**}  &  \epo  & Fluence, S$_\gamma$  & Fluence \\
 &  & error &  & error  & (keV)  & error  & ($10^{-7}~erg~cm^{-2}$)  & error \\
\hline
GRB080715950    & -1.10  & 0.08  & -2.30  & 0.35  & 230  & 42  & 50.4  & 0.8 \\
GRB080723557    & -0.76  & 0.03  & -2.15  & 0.04  & 200  & 8  & 722.3  & 2.5 \\
GRB080723985    & -0.93  & 0.03  & -2.30  &  & 431  & 22  & 307.9  & 2.1 \\
GRB080724401    & -0.91  & 0.09  & -2.10  & 0.08  & 104  & 10  & 156.9  & 0.5 \\
GRB080727964    & -0.94  & 0.09  & -2.49  & 0.36  & 180  & 24  & 133.0  & 0.8 \\
\hline
GRB080730786    & -0.75  & 0.06  & -2.30  &  & 135  & 6  & 63.5  & 0.8 \\
GRB080803772    & -0.47  & 0.13  & -2.30  &  & 298  & 45  & 43.9  & 0.7 \\
GRB080804456    & -1.09  & 0.13  & -2.30  &  & 196  & 43  & 80.0  & 0.9 \\
GRB080805496    & -1.82  & 0.14  & -2.30  &  & 31  & 12  & 17.5  & 0.5 \\
GRB080806896    & -0.66  & 0.15  & -2.34  & 0.08  & 47  & 3  & 132.8  & 1.9 \\
\hline
\textbf{GRB080807993}    & -0.92  & 0.06  & -2.30  &  & 618  & 118  & 73.0  & 0.9 \\
GRB080808565    & -1.11  & 0.10  & -2.44  & 0.19  & 80  & 8  & 39.7  & 0.4 \\
GRB080809808    & -1.26  & 0.31  & -2.20  & 0.22  & 63  & 21  & 41.4  & 0.5 \\
GRB080812889    & -0.18  & 0.30  & -2.18  & 0.31  & 145  & 29  & 24.5  & 0.4 \\
GRB080816503    & -0.87  & 0.05  & -2.61  & 0.20  & 126  & 7  & 133.0  & 0.8 \\
\hline
\textbf{GRB080816989}    & -0.54  & 0.10  & -2.30  &  & 1561  & 285  & 33.0  & 0.9 \\
GRB080817161    & -0.93  & 0.02  & -2.10  & 0.06  & 354  & 19  & 532.5  & 0.7 \\
GRB080818579    & -1.47  & 0.09  & -2.30  &  & 236  & 83  & 38.0  & 0.6 \\
GRB080818945    & -1.43  & 0.13  & -2.30  &  & 68  & 9  & 17.4  & 0.2 \\
GRB080821332    & -0.76  & 0.11  & -2.58  & 0.30  & 110  & 11  & 35.9  & 0.2 \\
\hline
GRB080824909    & -1.11  & 0.11  & -2.25  & 0.29  & 143  & 29  & 27.3  & 0.6 \\
GRB080825593    & -0.65  & 0.03  & -2.30  & 0.07  & 174  & 6  & 341.9  & 1.0 \\
GRB080829790    & -0.83  & 0.13  & -2.30  &  & 75  & 5  & 25.2  & 0.2 \\
GRB080830368    & -1.05  & 0.08  & -2.30  &  & 269  & 55  & 70.0  & 1.1 \\
GRB080904886    & -0.76  & 0.24  & -2.51  & 0.08  & 31  & 2  & 52.4  & 0.7 \\
\hline
GRB080905570    & -0.99  & 0.16  & -2.74  & 0.48  & 74  & 9  & 40.9  & 0.6 \\
GRB080906212    & -0.52  & 0.10  & -2.06  & 0.09  & 125  & 11  & 58.7  & 1.4 \\
GRB080913735    & -0.27  & 0.18  & -2.30  &  & 102  & 8  & 35.4  & 0.9 \\
GRB080924766    & -1.20  & 0.18  & -2.05  & 0.14  & 84  & 23  & 47.3  & 0.8 \\
GRB080925775    & -1.08  & 0.04  & -2.09  & 0.08  & 163  & 14  & 184.7  & 0.4 \\
\hline
\textbf{GRB081006604}    & -0.32  & 0.37  & -1.99  & 0.40  & 538  & 247  & 8.3  & 0.2 \\
GRB081009140    & -1.53  & 0.02  & -4.12  & 0.45  & 31  & 0  & 383.0  & 0.5 \\
GRB081012549    & -0.10  & 0.23  & -2.30  &  & 326  & 57  & 45.1  & 1.1 \\
GRB081017474    & -0.85  & 0.31  & -2.30  &  & 66  & 10  & 13.9  & 0.2 \\
GRB081021398    & 0.19  & 0.24  & -2.78  & 0.42  & 127  & 12  & 57.4  & 0.8 \\
\hline
GRB081024851    & -1.11  & 0.09  & -2.30  &  & 83  & 7  & 62.7  & 0.7 \\
GRB081025349    & -0.44  & 0.14  & -2.21  & 0.33  & 255  & 41  & 63.2  & 1.2 \\
GRB081028538    & -0.75  & 0.17  & -3.08  & 0.65  & 67  & 6  & 22.7  & 0.3 \\
GRB081101532    & -0.72  & 0.04  & -2.30  &  & 524  & 33  & 150.8  & 3.5 \\
GRB081102739    & -0.90  & 0.11  & -2.30  &  & 134  & 15  & 37.6  & 0.9 \\
\hline
GRB081118876    & -0.75  & 0.13  & -2.23  & 0.09  & 64  & 6  & 49.4  & 0.4 \\
GRB081122520    & -0.80  & 0.07  & -2.51  & 0.35  & 197  & 20  & 75.4  & 0.8 \\
GRB081125496    & -0.51  & 0.06  & -2.37  & 0.11  & 161  & 9  & 185.4  & 1.3 \\
GRB081126899    & -0.78  & 0.05  & -2.30  &  & 325  & 26  & 114.5  & 0.7 \\
GRB081129161    & -1.04  & 0.06  & -2.28  & 0.24  & 259  & 34  & 161.9  & 1.5 \\
\hline
GRB081130629    & -0.98  & 0.11  & -2.30  &  & 174  & 26  & 32.2  & 0.6 \\
GRB081206275    & 0.13  & 0.37  & -2.20  & 0.40  & 151 $^{ca}$  & 29  & 38.6  & 0.6 \\
\textbf{GRB081206987}    & -1.04  & 0.22  & -2.30  &  & 262  & 113  & 11.3  & 0.3 \\
GRB081207680    & -0.59  & 0.03  & -2.08  & 0.08  & 416  & 24  & 486.2  & 1.0 \\
GRB081215784    & -0.78  & 0.01  & -2.29  & 0.05  & 442  & 13  & 546.7  & 0.6 \\
\hline
GRB081215880    & -0.29  & 0.28  & -2.59  & 0.38  & 116  & 16  & 17.8  & 0.4 \\
GRB081217983    & -0.73  & 0.07  & -2.45  & 0.26  & 163  & 15  & 96.2  & 1.4 \\
GRB081224887    & -0.77  & 0.02  & -2.30  &  & 401  & 13  & 375.7  & 1.7 \\
GRB081226156    & -1.30  & 0.39  & -2.05  & 0.07  & 36  & 13  & 39.5  & 0.2 \\
GRB081231140    & -1.17  & 0.04  & -1.96  & 0.07  & 242  & 31  & 161.2  & 1.2 \\
\hline
\end{tabular}\\
\tablefoot{
\tablefoottext{**}{references for the spectral parameters: $^{ca)}$ \cite{Vonkienlin2008} }\\
}
\end{table*}

\begin{table*}
\addtocounter{table}{-1}
\caption{Continued}
\centering
\begin{tabular}{|lcccccccc|}
\hline
GRB Name  & $\alpha$  & $\alpha$ & $\beta$\tablefootmark{a}  & $\beta$ & \epo  \tablefootmark{**}  &  \epo  & Fluence  & Fluence \\
 &  & error &  & error  & (keV)  & error  & ($10^{-7}~erg~cm^{-2}$)  & error \\
\hline
GRB090101758    & -1.10  & 0.05  & -2.30  &  & 107  & 6  & 122.9  & 1.1 \\
GRB090112332    & -1.41  & 0.07  & -2.30  &  & 276  & 73  & 39.2  & 0.7 \\
GRB090112729    & -0.75  & 0.06  & -2.43  & 0.14  & 139 $^{cb}$  & 9  & 92.3  & 1.1 \\
GRB090113778    & -1.20  & 0.19  & -2.06  & 0.29  & 137  & 58  & 15.7  & 0.5 \\
GRB090117632    & -1.37  & 0.06  & -2.04  & 0.25  & 350  & 117  & 119.4  & 2.0 \\
\hline
GRB090126227    & -1.27  & 0.20  & -2.30  &  & 44  & 5  & 11.0  & 0.2 \\
GRB090129880    & -1.49  & 0.08  & -2.12  & 0.23  & 171  & 53  & 55.7  & 0.6 \\
GRB090131090    & -1.18  & 0.07  & -2.19  & 0.04  & 54  & 4  & 175.2  & 0.7 \\
\textbf{GRB090202347}    & -1.31  & 0.06  & -2.30  &   & 570 $^{cc}$  & 170  & 49.5  & 0.3 \\
\textbf{GRB090207777}    & -1.19  & 0.10  & -2.30  &  & 371  & 108  & 24.1  & 0.4 \\
\hline
GRB090217206    & -0.94  & 0.02  & -2.30  &  & 673  & 41  & 275.2  & 0.3 \\
GRB090222179    & -0.97  & 0.14  & -2.56  & 0.66  & 148 $^{cd}$  & 27  & 32.3  & 0.5 \\
\textbf{GRB090227310}    & -0.92  & 0.07  & -1.92  & 0.16  & 1678  & 565  & 28.6  & 0.2 \\
GRB090228976    & -0.76  & 0.26  & -2.30  &  & 142  & 34  & 9.6  & 0.7 \\
\textbf{GRB090301315}    & -1.05  & 0.13  & -2.09  & 0.47  & 480  & 213  & 22.7  & 0.4 \\
\hline
GRB090306245    & -0.85  & 0.23  & -2.30  &  & 97  & 16  & 13.7  & 0.4 \\
GRB090307167    & -0.63  & 0.37  & -2.30  &  & 153  & 47  & 10.9  & 0.4 \\
GRB090310189    & -0.87  & 0.11  & -2.84  & 0.89  & 123  & 15  & 55.4  & 0.6 \\
GRB090316311    & -1.15  & 0.18  & -2.30  &  & 136  & 34  & 10.6  & 0.2 \\
GRB090319622    & -0.84  & 0.14  & -2.26  & 0.28  & 152  & 27  & 60.3  & 0.6 \\
\hline
GRB090327404    & -0.42  & 0.18  & -2.47  & 0.21  & 84  & 8  & 28.2  & 0.6 \\
GRB090330279    & -0.81  & 0.07  & -2.04  & 0.11  & 203  & 23  & 118.1  & 0.4 \\
GRB090403314    & -1.11  & 0.19  & -2.30  &  & 222  & 85  & 10.9  & 0.2 \\
GRB090411991    & -0.74  & 0.12  & -1.98  & 0.11  & 172  & 29  & 62.1  & 0.9 \\
GRB090413122    & -0.83  & 0.18  & -1.99  & 0.33  & 304  & 108  & 32.3  & 0.5 \\
\hline
GRB090419997    & -1.38  & 0.05  & -2.30  &  & 150  & 14  & 95.4  & 2.4 \\
GRB090425377    & -1.29  & 0.11  & -2.03  & 0.05  & 69 $^{ce}$  & 11  & 181.3  & 1.5 \\
GRB090426690    & -1.26  & 0.10  & -2.10  & 0.38  & 281  & 107  & 35.4  & 0.9 \\
GRB090428441    & -0.46  & 0.22  & -2.30  &  & 84  & 8  & 10.4  & 0.6 \\
GRB090502777    & -0.78  & 0.29  & -2.36  & 0.14  & 48  & 7  & 35.0  & 0.3 \\
\hline
GRB090509215    & -0.89  & 0.17  & -2.42  & 0.80  & 206  & 57  & 54.2  & 0.7 \\
GRB090511684    & -0.77  & 0.28  & -1.93  & 0.29  & 167  & 72  & 24.9  & 0.8 \\
\textbf{GRB090513916}    & -0.87  & 0.11  & -2.30  &  & 813  & 241  & 49.4  & 1.8 \\
GRB090514726    & -0.57  & 0.20  & -2.13  & 0.26  & 189  & 41  & 22.5  & 0.3 \\
GRB090514734    & -1.23  & 0.14  & -2.24  & 0.18  & 85  & 16  & 95.5  & 2.1 \\
\hline
GRB090516137    & -0.99  & 0.04  & -2.12  & 0.17  & 328  & 39  & 168.4  & 1.8 \\
GRB090518244    & -0.64  & 0.17  & -2.74  & 0.53  & 105  & 13  & 21.1  & 0.7 \\
GRB090519462    & -0.59  & 0.39  & -2.30  &  & 98  & 24  & 43.8  & 0.5 \\
GRB090520850    & -0.71  & 0.10  & -2.30  &  & 215  & 24  & 33.2  & 1.0 \\
GRB090520876    & -0.92  & 0.13  & -2.76  & 0.15  & 41  & 2  & 61.8  & 0.4 \\
\hline
GRB090524346    & -1.00  & 0.05  & -2.68  & 0.20  & 95  & 5  & 165.9  & 0.6 \\
GRB090528173    & -0.55  & 0.24  & -2.24  & 0.06  & 40  & 4  & 65.6  & 1.1 \\
GRB090528516    & -1.23  & 0.02  & -2.14  & 0.08  & 229  & 17  & 435.2  & 0.9 \\
GRB090529564    & -1.18  & 0.06  & -1.91  & 0.08  & 205  & 38  & 86.9  & 0.3 \\
GRB090530760    & -0.98  & 0.03  & -2.62  & 0.05  & 64  & 1  & 598.9  & 1.6 \\
\hline
\textbf{GRB090602564}    & -0.71  & 0.13  & -2.36  & 0.47  & 715  & 197  & 27.9  & 0.6 \\
\textbf{GRB090610648}    & -0.79  & 0.10  & -2.30  &  & 1905  & 681  & 13.5  & 0.6 \\
GRB090610723    & -0.64  & 0.22  & -2.63  & 0.61  & 115  & 19  & 39.6  & 0.6 \\
GRB090612619    & -0.74  & 0.08  & -2.37  & 0.45  & 468  & 77  & 58.2  & 0.7 \\
GRB090620400    & -0.43  & 0.05  & -2.72  & 0.18  & 160  & 6  & 132.6  & 0.4 \\
\hline
GRB090621185    & -1.10  & 0.20  & -2.12  & 0.09  & 56 $^{cf}$  & 11  & 108.4  & 2.1 \\
GRB090623107    & -0.77  & 0.05  & -2.05  & 0.14  & 367  & 42  & 118.4  & 0.7 \\
GRB090623913    & -1.28  & 0.08  & -2.30  &  & 221  & 48  & 21.6  & 0.6 \\
GRB090625560    & -0.78  & 0.21  & -2.30  &  & 147  & 27  & 24.6  & 0.9 \\
GRB090626189    & -1.29  & 0.02  & -1.98  & 0.02  & 175 $^{cg}$  & 12  & 630.5  & 1.1 \\
\hline
\end{tabular}\\
\tablefoot{
\tablefoottext{**}{references for the spectral parameters: $^{cb)}$ \cite{Vanderhorst2009} -- $^{cc)}$ \cite{Vonkienlin2009a} -- $^{cd)}$ \cite{Vonkienlin2009b} -- $^{ce)}$ \cite{Vonkienlin2009c} -- $^{cf)}$ \cite{Rau2009} -- $^{cg)}$ \cite{Vonkienlin2009d} }\\
}
\end{table*}

\begin{table*}
\addtocounter{table}{-1}
\caption{Continued.}
\centering
\begin{tabular}{|lcccccccc|}
\hline
GRB Name  & $\alpha$  & $\alpha$ & $\beta$\tablefootmark{a}  & $\beta$ & \epo  \tablefootmark{**}  &  \epo  & Fluence  & Fluence \\
 &  & error &  & error  & (keV)  & error  & ($10^{-7}~erg~cm^{-2}$)  & error \\
\hline
GRB090630311    & -1.36  & 0.14  & -2.30  &  & 63  & 7  & 10.8  & 0.1 \\
GRB090703329    & -1.19  & 0.31  & -2.19  & 0.38  & 87  & 38  & 8.5  & 0.3 \\
GRB090704242    & -1.25  & 0.06  & -2.30  &  & 275  & 47  & 84.8  & 1.0 \\
GRB090709630    & -0.49  & 0.23  & -1.96  & 0.13  & 106  & 20  & 22.1  & 0.4 \\
GRB090713020    & -0.13  & 0.11  & -3.07  & 0.31  & 92  & 4  & 94.8  & 0.4 \\
\hline
GRB090717034    & -0.94  & 0.06  & -1.95  & 0.04  & 108  & 8  & 232.3  & 0.7 \\
GRB090718762    & -1.11  & 0.03  & -2.47  & 0.15  & 162  & 9  & 250.2  & 1.2 \\
GRB090719063    & -0.72  & 0.02  & -3.09  & 0.27  & 250  & 7  & 468.5  & 1.6 \\
GRB090720276    & -0.59  & 0.12  & -2.30  &  & 105  & 8  & 32.2  & 0.3 \\
\textbf{GRB090720710}    & -1.04  & 0.04  & -2.42  & 0.35  & 1242  & 254  & 142.4  & 0.2 \\
\hline
GRB090730608    & -0.50  & 0.20  & -2.34  & 0.35  & 137  & 22  & 31.8  & 0.8 \\
GRB090804940    & -0.58  & 0.04  & -3.62  & 0.35  & 95  & 2  & 144.3  & 1.9 \\
GRB090805622    & -1.00  & 0.18  & -2.30  &  & 86  & 10  & 57.9  & 0.5 \\
GRB090809978    & -0.81  & 0.04  & -2.06  & 0.06  & 187  & 12  & 216.3  & 1.3 \\
GRB090810659    & -0.84  & 0.23  & -2.36  & 0.08  & 38  & 4  & 98.9  & 0.9 \\
\hline
GRB090810781    & -0.44  & 0.34  & -2.10  & 0.07  & 47  & 7  & 51.5  & 0.6 \\
\textbf{GRB090811696}    & -0.50  & 0.29  & -2.30  &  & 685  & 231  & 10.5  & 0.2 \\
GRB090813174    & -1.25  & 0.14  & -2.14  & 0.12  & 81  & 17  & 33.3  & 0.4 \\
GRB090814950    & -0.84  & 0.08  & -2.22  & 0.35  & 340  & 60  & 159.8  & 3.9 \\
GRB090815300    & -0.40  & 0.39  & -2.30  &  & 165  & 43  & 14.3  & 0.4 \\
\hline
GRB090815438    & -1.05  & 0.17  & -2.83  & 0.21  & 34  & 2  & 48.9  & 1.6 \\
GRB090815946    & -1.00  & 0.16  & -2.30  &  & 154  & 42  & 28.8  & 0.3 \\
GRB090820027    & -0.68  & 0.01  & -2.60  & 0.04  & 209  & 3  & 1536.0  & 1.8 \\
GRB090826068    & -0.68  & 0.21  & -2.30  &  & 163  & 35  & 8.5  & 0.4 \\
GRB090829672    & -1.44  & 0.04  & -2.10  & 0.10  & 183 $^{ch}$  & 31  & 766.4  & 1.6 \\
\hline
GRB090829702    & -0.54  & 0.15  & -2.14  & 0.17  & 121  & 16  & 48.1  & 0.6 \\
GRB090831317    & -1.43  & 0.05  & -2.20  & 0.21  & 190  & 38  & 94.4  & 0.7 \\
\textbf{GRB090902401}    & 0.06  & 0.20  & -2.30  &  & 428  & 54  & 16.7  & 0.4 \\
GRB090904058    & -1.13  & 0.07  & -2.30  & 0.12  & 113  & 11  & 217.1  & 2.2 \\
\textbf{GRB090904581}    & -0.68  & 0.20  & -2.37  & 0.97  & 840  & 401  & 16.4  & 0.3 \\
\hline
\textbf{GRB090908341}    & -0.98  & 0.06  & -2.41  & 0.64  & 3483  & 1212  & 26.0  & 0.2 \\
GRB090909487    & -1.26  & 0.23  & -2.30  &  & 205  & 88  & 57.3  & 2.0 \\
GRB090910812    & -0.87  & 0.05  & -2.77  & 0.62  & 284  & 28  & 186.6  & 2.1 \\
GRB090912660    & -0.38  & 0.23  & -2.56  & 0.34  & 92  & 11  & 103.8  & 1.7 \\
GRB090915650    & -0.78  & 0.22  & -1.91  & 0.21  & 197  & 66  & 29.9  & 0.4 \\
\hline
\textbf{GRB090917661}    & -0.77  & 0.24  & -2.30  &  & 231  & 62  & 10.8  & 0.4 \\
GRB090922539    & -0.93  & 0.05  & -2.50  & 0.18  & 142  & 9  & 110.5  & 0.5 \\
GRB090922605    & -0.94  & 0.23  & -1.86  & 0.13  & 306  & 133  & 45.1  & 1.0 \\
GRB090925389    & -0.40  & 0.19  & -2.01  & 0.09  & 101  & 13  & 89.1  & 3.1 \\
GRB090928646    & -0.56  & 0.25  & -1.97  & 0.22  & 233  & 70  & 19.5  & 0.7 \\
\hline
GRB090929190    & -0.48  & 0.07  & -2.30  &  & 527  & 40  & 81.8  & 1.0 \\
GRB091005679    & -0.76  & 0.28  & -2.16  & 0.49  & 161  & 57  & 14.1  & 0.7 \\
GRB091010113    & -1.10  & 0.03  & -2.30  &  & 148  & 6  & 99.6  & 0.6 \\
\textbf{GRB091020977}    & -0.99  & 0.05  & -2.73  & 0.62  & 1201  & 241  & 107.5  & 0.6 \\
GRB091024380    & -0.88  & 0.06  & -2.70  & 0.50  & 190  & 18  & 255.3  & 0.5 \\
\hline
GRB091030613    & -0.66  & 0.13  & -2.56  & 0.39  & 149  & 19  & 44.8  & 0.4 \\
GRB091030828    & -0.96  & 0.03  & -2.50  & 0.29  & 513  & 41  & 296.3  & 2.0 \\
GRB091031500    & -0.86  & 0.04  & -2.10  & 0.13  & 461  & 50  & 152.9  & 0.9 \\
GRB091101143    & -0.78  & 0.08  & -2.40  & 0.20  & 127  & 11  & 78.4  & 0.8 \\
GRB091103912    & -1.04  & 0.09  & -2.11  & 0.18  & 196  & 35  & 56.0  & 1.1 \\
\hline
GRB091107635    & -1.03  & 0.28  & -2.30  &  & 165  & 62  & 9.3  & 0.4 \\
GRB091109895    & -1.11  & 0.16  & -2.30  &  & 102  & 17  & 20.2  & 0.4 \\
GRB091112737    & -1.10  & 0.06  & -2.22  & 0.40  & 683  & 169  & 99.0  & 0.9 \\
\textbf{GRB091115177}    & -0.87  & 0.12  & -2.03  & 0.41  & 577  & 223  & 15.4  & 0.6 \\
GRB091117080    & -0.99  & 0.32  & -2.19  & 0.25  & 75  & 23  & 36.8  & 0.4 \\
\hline
\end{tabular}\\
\tablefoot{
\tablefoottext{**}{references for the spectral parameters: $^{ch)}$ \cite{Wilsonhodge2009} }\\
}
\end{table*}

\begin{table*}
\addtocounter{table}{-1}
\caption{Continued.}
\centering
\begin{tabular}{|lcccccccc|}
\hline
GRB Name  & $\alpha$  & $\alpha$ & $\beta$\tablefootmark{a}  & $\beta$ & \epo  &  \epo  & Fluence  & Fluence \\
 &  & error &  & error  & (keV)  & error  & ($10^{-7}~erg~cm^{-2}$)  & error \\
\hline
GRB091120191    & -1.07  & 0.03  & -3.22  & 0.45  & 124  & 4  & 285.3  & 4.1 \\
GRB091123081    & -0.92  & 0.31  & -2.24  & 0.33  & 88  & 26  & 21.3  & 0.9 \\
GRB091123298    & -0.81  & 0.08  & -2.18  & 0.12  & 157  & 15  & 640.4  & 3.7 \\
GRB091128285    & -1.01  & 0.03  & -2.30  &  & 193  & 9  & 403.9  & 4.5 \\
GRB091202072    & -0.77  & 0.36  & -2.32  & 0.43  & 94  & 27  & 16.7  & 0.3 \\
\hline
GRB091202219    & -0.95  & 0.19  & -1.96  & 0.11  & 101  & 26  & 68.0  & 1.5 \\
GRB091207333    & -0.72  & 0.13  & -2.11  & 0.18  & 152  & 23  & 53.7  & 1.1 \\
GRB091209001    & 0.29  & 0.21  & -3.04  & 0.26  & 62  & 3  & 100.3  & 1.9 \\
GRB091220442    & -1.00  & 0.11  & -2.31  & 0.15  & 86  & 10  & 58.3  & 0.5 \\
GRB091221870    & -0.63  & 0.09  & -2.04  & 0.09  & 165  & 17  & 89.4  & 2.2 \\
\hline
GRB091223511    & 0.34  & 0.24  & -2.19  & 0.16  & 130  & 13  & 86.9  & 0.5 \\
GRB091227294    & -0.97  & 0.06  & -2.30  &  & 287  & 33  & 68.9  & 1.1 \\
GRB091230712    & -0.62  & 0.22  & -2.30  &  & 100  & 13  & 25.8  & 0.9 \\
GRB091231206    & -0.72  & 0.09  & -2.64  & 0.37  & 151  & 15  & 97.6  & 2.1 \\
GRB100111176    & -1.40  & 0.16  & -2.30  &  & 202  & 98  & 11.5  & 0.2 \\
\hline
GRB100116897    & -1.03  & 0.02  & -2.30  &  & 1075  & 84  & 333.9  & 1.6 \\
\textbf{GRB100118100}    & -0.29  & 0.23  & -2.55  & 0.76  & 839  & 231  & 14.4  & 1.1 \\
GRB100122616    & -0.99  & 0.13  & -2.31  & 0.05  & 43  & 3  & 120.0  & 1.6 \\
GRB100130729    & -0.99  & 0.14  & -2.10  & 0.16  & 153  & 34  & 85.7  & 0.9 \\
GRB100130777    & -1.02  & 0.10  & -1.99  & 0.12  & 157  & 28  & 138.7  & 1.7 \\
\hline
GRB100131730    & -0.87  & 0.06  & -2.36  & 0.15  & 146  & 11  & 73.4  & 0.8 \\
GRB100201588    & -1.37  & 0.08  & -2.13  & 0.13  & 107  & 19  & 107.2  & 0.6 \\
GRB100204024    & -0.87  & 0.06  & -2.30  &  & 188  & 15  & 103.4  & 1.5 \\
GRB100204566    & -1.10  & 0.17  & -2.69  & 0.87  & 125  & 28  & 37.8  & 0.5 \\
GRB100205490    & -0.55  & 0.22  & -2.75  & 0.85  & 127  & 22  & 13.6  & 0.3 \\
\hline
\textbf{GRB100207665}    & -0.41  & 0.32  & -2.30  &  & 357  & 101  & 20.8  & 0.4 \\
GRB100211440    & -0.84  & 0.07  & -2.77  & 0.23  & 111  & 7  & 151.8  & 1.7 \\
GRB100212550    & 0.31  & 0.22  & -2.41  & 0.32  & 283  & 33  & 36.0  & 0.9 \\
GRB100212588    & -0.88  & 0.34  & -2.30  &  & 104  & 24  & 3.5  & 0.2 \\
GRB100218194    & -0.76  & 0.21  & -2.30  &  & 181  & 42  & 26.4  & 1.0 \\
\hline
\textbf{GRB100219026}    & -0.69  & 0.15  & -2.46  & 0.77  & 777  & 266  & 34.8  & 0.7 \\
GRB100224112    & -1.35  & 0.07  & -2.61  & 0.80  & 166  & 31  & 107.2  & 3.7 \\
GRB100225115    & -0.73  & 0.08  & -2.67  & 0.88  & 404  & 62  & 58.5  & 0.8 \\
GRB100225580    & -1.03  & 0.07  & -2.15  & 0.22  & 267  & 45  & 64.0  & 1.1 \\
\textbf{GRB100225703}    & -0.56  & 0.18  & -2.06  & 0.26  & 497  & 138  & 16.1  & 0.4 \\
\hline
GRB100301223    & -0.28  & 0.16  & -2.30  &  & 111  & 8  & 24.0  & 0.6 \\
\textbf{GRB100304004}    & -0.87  & 0.09  & -2.30  &  & 616  & 130  & 63.1  & 1.4 \\
GRB100311518    & -0.44  & 0.18  & -2.30  &  & 321  & 66  & 25.7  & 1.0 \\
GRB100313288    & -0.43  & 0.13  & -2.24  & 0.24  & 185  & 24  & 44.0  & 0.8 \\
GRB100322045    & -0.81  & 0.02  & -1.94  & 0.03  & 253  & 11  & 570.9  & 2.1 \\
\hline
\textbf{GRB100323542}    & -1.00  & 0.10  & -2.30  &  & 336  & 70  & 20.4  & 1.3 \\
GRB100324172    & -0.59  & 0.02  & -2.30  &  & 446  & 14  & 427.9  & 1.7 \\
GRB100325275    & -0.41  & 0.14  & -2.88  & 0.74  & 156  & 17  & 33.5  & 0.4 \\
GRB100326402    & -1.02  & 0.08  & -2.07  & 0.18  & 234  & 42  & 118.2  & 2.7 \\
GRB100330309    & -0.93  & 0.11  & -2.41  & 0.23  & 103  & 12  & 43.0  & 0.5 \\
\hline
\textbf{GRB100410740}    & -0.93  & 0.08  & -2.30  &  & 997  & 298  & 62.1  & 3.1 \\
\textbf{GRB100413732}    & -1.22  & 0.05  & -2.30  &  & 959  & 331  & 104.7  & 0.8 \\
GRB100420008    & -0.67  & 0.11  & -2.93  & 0.61  & 124  & 11  & 43.1  & 0.4 \\
\textbf{GRB100423244}    & -0.88  & 0.05  & -2.16  & 0.22  & 1215  & 260  & 79.2  & 1.2 \\
GRB100424729    & -0.60  & 0.14  & -2.59  & 0.73  & 273  & 45  & 74.1  & 0.6 \\
\hline
GRB100424876    & -0.96  & 0.05  & -2.30  &  & 233  & 20  & 149.5  & 1.7 \\
GRB100427356    & -0.78  & 0.15  & -2.93  & 0.83  & 104  & 13  & 22.8  & 0.6 \\
GRB100429999    & -0.79  & 0.17  & -2.32  & 0.54  & 215  & 56  & 27.8  & 0.4 \\
GRB100502356    & -1.19  & 0.04  & -2.30  &  & 508  & 80  & 156.0  & 2.1 \\
GRB100503554    & -0.99  & 0.05  & -2.30  &  & 241  & 21  & 173.2  & 4.1 \\
\hline
GRB100504806    & -0.59  & 0.19  & -2.30  &  & 82  & 7  & 23.3  & 1.3 \\
GRB100506653    & -1.08  & 0.15  & -2.00  & 0.28  & 232  & 90  & 24.2  & 0.5 \\
GRB100507577    & 0.46  & 0.11  & -3.48  & 0.55  & 134  & 5  & 99.7  & 1.2 \\
GRB100510810    & -0.79  & 0.20  & -2.90  & 0.53  & 65  & 7  & 37.2  & 0.5 \\
\textbf{GRB100511035}    & -1.38  & 0.02  & -2.25  & 0.26  & 1325  & 304  & 299.7  & 1.0 \\
\hline
\end{tabular}
\end{table*}

\begin{table*}
\addtocounter{table}{-1}
\caption{Continued.}
\centering
\begin{tabular}{|lcccccccc|}
\hline
GRB Name  & $\alpha$  & $\alpha$ & $\beta$\tablefootmark{a}  & $\beta$ & \epo  \tablefootmark{**}  &  \epo  & Fluence  & Fluence \\
 &  & error &  & error  & (keV)  & error  & ($10^{-7}~erg~cm^{-2}$)  & error \\
\hline
GRB100513879    & -1.34  & 0.06  & -2.30  &  & 238  & 38  & 37.1  & 0.5 \\
GRB100515467    & -1.07  & 0.08  & -2.22  & 0.17  & 145  & 19  & 61.1  & 0.5 \\
GRB100517072    & -1.00  & 0.29  & -1.93  & 0.05  & 45  & 11  & 65.9  & 0.2 \\
GRB100517154    & -1.11  & 0.13  & -2.30  &  & 48  & 3  & 27.9  & 0.3 \\
GRB100517243    & -1.26  & 0.37  & -2.66  & 0.39  & 39  & 6  & 26.9  & 0.4 \\
\hline
GRB100517639    & -0.36  & 0.09  & -2.30  &  & 87  & 3  & 29.1  & 1.3 \\
GRB100519204    & -0.85  & 0.04  & -2.30  &  & 108  & 3  & 206.8  & 2.3 \\
GRB100522157    & -1.39  & 0.09  & -2.30  &  & 158  & 27  & 38.6  & 0.4 \\
GRB100530737    & -1.16  & 0.28  & -2.31  & 0.73  & 107  & 47  & 4.8  & 0.2 \\
GRB100604287    & -1.08  & 0.12  & -2.20  & 0.23  & 167  & 34  & 55.1  & 0.4 \\
\hline
GRB100609783    & 0.32  & 0.10  & -2.30  &  & 152  & 5  & 174.2  & 6.1 \\
GRB100612726    & -0.73  & 0.04  & -2.49  & 0.10  & 113  & 4  & 136.3  & 3.6 \\
GRB100614498    & -0.91  & 0.22  & -2.21  & 0.08  & 41  & 5  & 196.3  & 3.3 \\
GRB100619015    & -1.30  & 0.07  & -2.30  &  & 120  & 12  & 112.9  & 0.7 \\
GRB100701490    & -0.85  & 0.03  & -2.24  & 0.10  & 1132  & 114  & 260.3  & 0.4 \\
\hline
GRB100704149    & -0.93  & 0.09  & -2.00  & 0.14  & 175  & 30  & 104.0  & 1.1 \\
GRB100707032    & -0.91  & 0.02  & -2.30  &   & 283 $^{ci}$  & 7  & 877.2  & 1.6 \\
GRB100709602    & -1.07  & 0.08  & -2.30  &  & 183  & 27  & 80.8  & 0.8 \\
\hline
\end{tabular}\\
\tablefoot{
\tablefoottext{**}{references for the spectral parameters: $^{ci)}$ \cite{Wilsonhodge2009} }\\
}
\end{table*}

\end{appendix}

\end{document}